%% file: ms.tex
\documentclass[numberedappendix]{emulateapj}

\shorttitle{Cosmological constraints from gravitational lens \ourlens}
\shortauthors{Suyu et al.}

\usepackage{natbib}
\input{ms_macro}

\begin{document}

\title{Dissecting the Gravitational Lens \ourlens. 
II. Precision Measurements of the Hubble Constant, Spatial Curvature, and the
Dark Energy Equation of State\altaffilmark{*}}

\author{S.~H.~Suyu\altaffilmark{1},
  P.~J.~Marshall\altaffilmark{2,3},
  M.~W.~Auger\altaffilmark{3,4},
  S.~Hilbert\altaffilmark{1,5},
  R.~D.~Blandford\altaffilmark{2},
  L.~V.~E.~Koopmans\altaffilmark{6},
  C.~D.~Fassnacht\altaffilmark{4}, and
  T.~Treu\altaffilmark{3,7}}

\altaffiltext{*}{Based in part on observations made with the NASA/ESA \textit{Hubble Space Telescope}, obtained at the Space Telescope Science Institute, which is operated by the Association of Universities for Research in Astronomy, Inc., under 
NASA contract NAS 5-26555. These observations are associated with program 
GO-10158.}
\altaffiltext{1}{Argelander-Institut f{\"u}r Astronomie, Auf dem H{\"u}gel 71, 53121 Bonn, Germany}
\altaffiltext{2}{Kavli Institute for Particle Astrophysics and Cosmology, Stanford University, PO Box 20450, MS 29, Stanford, CA 94309, USA}
\altaffiltext{3}{Department of Physics, University of California, Santa Barbara, CA 93106-9530, USA}
\altaffiltext{4}{Department of Physics, University of California at Davis, 1 Shields Avenue, Davis, CA 95616, USA}
\altaffiltext{5}{Max-Planck-Institut f{\"u}r Astrophysik, Karl-Schwarzschild-Str. 1, 85741 Garching, Germany}
\altaffiltext{6}{Kapteyn Astronomical Institute, P.O.~Box 800, 9700AV Groningen, The Netherlands}
\altaffiltext{7}{Sloan Fellow, Packard Fellow} 
\email{suyu@astro.uni-bonn.de}


\begin{abstract}

  Strong gravitational lens systems with measured time delays between the
  multiple images provide a method for measuring the ``time-delay distance''
  to the lens, and thus the Hubble constant.  We
  present a Bayesian analysis of the strong gravitational lens system
  \ourlens, incorporating (i) new, deep {\it Hubble Space Telescope} (\HST)
  observations, (ii) a new velocity dispersion measurement of
  $260\pm15\kms$ for the primary lens
  galaxy, and (iii) an updated study of the lens' environment.  Our analysis of
  the \HST\ images takes into account the extended source surface brightness, 
  and the dust extinction and optical emission by the 
  interacting lens galaxies.  
  When modeling the stellar dynamics of the primary lens galaxy, 
  the lensing effect, and
  the environment of the lens, we explicitly include the total mass
  distribution profile logarithmic slope $\slope$ and the external
  convergence $\kext$; 
  we marginalize over these parameters, 
  assigning well-motivated priors for them, and so turn the major
  systematic errors into statistical ones.
  The \HST\ images provide one such prior, constraining 
  the lens mass density profile logarithmic 
  slope to be~$\slope = 2.08 \pm 0.03$; a combination of numerical
  simulations and photometric observations of the \ourlens\
  field provides an estimate of the prior for~$\kext$: 
  $\kextOBS^{\kextOBShiOne}_{\kextOBSloOne}$.
  This latter distribution 
  dominates the final uncertainty on $H_0$.
  Fixing the cosmological parameters at $\OM=0.3$, $\OL=0.7$, and 
  $w=-1$ in order to compare
  with previous work on this system, we find 
  $H_0 = \HK ^{+\HKerrOne}_{-\HKerrOne} \kmsMpc$.
  The new data provide an increase in precision of 
  more than a factor of two, even including the marginalization over~$\kext$.
  Relaxing the prior probability density function for the cosmological
  parameters to that derived from 
  the WMAP 5-year data set, we find that the \ourlens\ data set breaks the
  degeneracy between $\OM$ and $\OL$ at $w=-1$ and constrains the curvature
  parameter to be $\OkWOTwo$ (95\% CL),  
  a level of precision comparable
  to that afforded by the current Type Ia SNe sample. 
  Asserting a flat spatial geometry, we find that, 
  in combination with WMAP, $H_0 = \HWW^{\HWWhiOne}_{\HWWloOne} \kmsMpc$ and
  $w=\wWW^{\wWWhiOne}_{\wWWloOne}$ (68\% CL), suggesting that the 
  observations of \ourlens\ constrain~$w$ 
  as tightly as do the current Baryon Acoustic Oscillation data. 

\end{abstract}

\keywords{cosmology: observations --- distance scale --- galaxies:
  individual (B1608+656) --- gravitational lensing: strong ---
  methods: data analysis}


\section{Introduction}
\label{sec:intro}

The Hubble constant ($H_0$, measured in units of $\kmsMpc$) is one of the key
cosmological parameters since it sets
the present age, size, and critical density of the Universe.  

Methods for measuring
the Hubble constant include Type Ia supernovae (SNe Ia)
\citep[e.g.][]{Tammann79, RiessEtal09}, the Sunyaev-Zel'dovich
effect \citep[e.g.][]{SunyaevZeldovich80, BonamenteEtal06},
the expanding photosphere method for Type II supernovae
\citep[e.g.][]{KirshnerKwan74,SchmidtEtal94}, 
and maser distances \citep[e.g.][]{Herrnstein99, MacriEtal06}.
However, perhaps the two most well-known recent measurements come from 
the \textit{Hubble Space Telescope} (\HST) Key Project (KP)
\citep{FreedmanEtal01} and the Wilkinson Microwave Anisotropy Probe (WMAP)
observations of the cosmic microwave background (CMB)
\citep[e.g.][]{KomatsuEtal09}. The
\HST\ KP measurement of $H_0$ is based on secondary distance
indicators (including Type Ia supernovae, Tully-Fisher, surface
brightness fluctuations, Type II supernovae, and the fundamental
plane) that are calibrated using Cepheid distances to nearby galaxies
with a zero point in the Large Magellanic Cloud.  The resulting
Hubble constant is $72\pm8\,\kmsMpc$ \citep{FreedmanEtal01}.
We note that the largest contributor to the systematic error from the
distance ladder of which this measurement depends is the metallicity
dependence of the Cepheid period-luminosity relation.
More recently, \citet{RiessEtal09} addressed some of these systematic effects
with an improved differential distance ladder 
using Cepheids, SNe Ia, and the maser galaxy NGC 4258, finding 
$H_0=74.2 \pm 3.6 \kmsMpc$, a 5\% local 
measurement of Hubble's constant.

The five year measurement made using WMAP temperature and
polarization data is $H_0=71.9^{+2.6}_{-2.7} \kmsMpc$ \citep{DunkleyEtal09}, 
under the assumption that the Universe
is flat and that the dark energy is described by a cosmological constant
(with equation of state parameter $w=-1$).  
The uncertainty in $H_0$ increases markedly if either
of these two assumptions is relaxed, due to degeneracies
with other cosmological parameters.  For example, WMAP 
gives $H_0\sim50\kmsMpc$ without the flatness assumption, and 
$H_0 = 74^{+15}_{-14}\kmsMpc$ for a flat Universe with time-independent $w$ 
not fixed at $w=-1$.  
As $H_0$ is such
an important parameter, it is essential to measure it using multiple
methods.  
In this paper, we use a single strong gravitational
lens as an independent  
probe of $H_0$, and explore its systematic errors and relations with other
cosmological parameters to provide guidance for future studies.  We will show
that the single lens is competitive with those of the best current
cosmographic probes.  
Given the current progress in measuring time delays
\citep[e.g.,][]{VuissozEtal07, VuissozEtal08, ParaficzEtal09}, the
methodology in this paper should lead to substantial advances when
applied to samples of gravitational lenses.  

Strong gravitational lensing occurs when a source galaxy is lensed
into multiple images by a galaxy lying along its line of
sight.  The principle of using strong gravitational lens systems with
time-variable sources to measure the Hubble
constant is well understood (e.g. \citeauthor{Refsdal64}~\citeyear{Refsdal64}, 
\citeauthor*{SchneiderEtal06}~\citeyear{SchneiderEtal06}).
The relative time delays between the multiple images are inversely
proportional to $H_0$ via a combination of angular diameter distances
and depend on the lens potential (mass) distribution. We refer to the
combination of angular diameter distances as the ``time-delay
distance''.  By measuring the time delays and modeling the lens
potential, one can infer the value for the time-delay distance; this
distance-like quantity is primarily sensitive to $H_0$ but depends
also on other cosmological parameters which must be factored into the
analysis.  The direct measurement of the time-delay distance 
means that gravitational lensing is independent of distance ladders.
  
Despite being an elegant method, gravitational lensing has its
limitations.  Perhaps the most well-known is the ``mass-sheet
degeneracy'' between $H_0$ and external convergence
\citep*{FalcoEtal85}.
There is also a
degeneracy between $H_0$ and the slope of the lens mass distribution,
especially for lenses where the configuration is nearly symmetric 
\citep[e.g.][]{Wucknitz02}. In such
cases, the image positions are at approximately the same radial
distance from the lens center and so the slope is poorly
constrained.  
In both cases the remedy is to provide more information. Modeling the mass
environment of the lens can, in principle,
independently constrain the external convergence (e.g.,
\citeauthor{KeetonZabludoff04}~\citeyear{KeetonZabludoff04};
\citeauthor{FassnachtEtal06}~\citeyear{FassnachtEtal06}; 
Blandford et al.~in preparation); likewise, lens galaxy stellar velocity
dispersion measurements 
\citep[e.g.,][]{GroginNarayan96a, GroginNarayan96b, TonryFranx99,
    KoopmansTreu02, TreuKoopmans02, BarnabeKoopmans07, McKeanEtal09}
and analysis of any extended images 
\citep[e.g.,][]{DyeWarren05, DyeEtal08}
can constrain the
mass distribution slope. 

A measurement of $H_0$ to better than a few percent precision would
provide
the single most useful complement to results obtained from studies of
the CMB for dark energy studies \citep[e.g.][]{Hu05, RiessEtal09}.
Dark energy has been used to explain the accelerating Universe,
discovered using luminosity distances to SNe Ia \citep{RiessEtal98,
  PerlmutterEtal99}.  Efforts in studying dark energy often
characterize it by a constant equation of state parameter $w$ (where
$w=-1$ corresponds to a cosmological constant) and assume a flat
Universe.  These include \citet{PerlmutterEtal99}, who in their Figure
10 constrained $w\lesssim -0.65$ for present day matter density values
of $\OM\ge0.2$, and \citet{EisensteinEtal05}, who combined their
angular diameter distance measurement to $z=0.35$ from Baryon Acoustic
Oscillations (BAO) with WMAP data \citep{SpergelEtal07} to obtain
$w=-0.80\pm0.18$.  Recently, \citet{KomatsuEtal09} measured
$w=-0.992^{+0.061}_{-0.062}$ by combining WMAP 5-year results (WMAP5) with
observations of SNe Ia \citep{KowalskiEtal08} and BAO
\citep{PercivalEtal07}.  \citet{KomatsuEtal09} also explored more
general dark energy descriptions.  In our study, we combine the
time-delay distance measurement from \ourlens\ with WMAP data to derive
a constraint on $w$, and compare the constraining power of \ourlens\
to that of other cosmographic probes.

In this paper, we present an accurate measurement of $H_0$ from the
gravitational lens \ourlens.  A comprehensive lensing analysis of the
lens system is in a companion paper (\paperI; \citeauthor{SuyuEtal09}~\citeyear{SuyuEtal09}).  
Using the results from
Paper I, we focus in this paper on techniques required to break the 
mass-sheet degeneracy in order to infer a value of $H_0$ with well-understood
uncertainty. We then explore the influence of this measurement on other
cosmological parameters.  

The organization of the paper is as follows.  In Section
\ref{sec:H0theory}, we briefly review the theory behind using
gravitational lenses to measure $H_0$, include a description of the
mass-sheet degeneracy, and describe the dynamics modeling for the
measured velocity dispersion.  In Section \ref{sec:H0ProbTheory}, we
outline the probability theory for combining various data sets and for
including cosmological priors.  In Section \ref{sec:LensModel}, we
present the gravitational lens \ourlens\ as a candidate for measuring
$H_0$, and show the lens modeling results.  
We present the new velocity dispersion measurement and the
stellar dynamics modeling in Section \ref{sec:StellDyn}.  The study of
the convergence accumulated along the line of sight to \ourlens\ 
is discussed in Section
\ref{sec:LensEnv}.  The priors for our model parameters are described
in Section \ref{sec:Priors}. Finally, in Section \ref{sec:H0} we combine the
lensing, dynamics and external convergence analyses 
to break the mass-sheet
degeneracy and infer $H_0$ from the \ourlens\ data set.  We then
show how \ourlens\ aids in constraining flatness and 
measuring $w$ when combined with WMAP, before
concluding in Section~\ref{sec:conc}.

Throughout this paper, we assume a $w$-CDM universe where dark energy
is described by a time-independent equation of state with parameter 
$w=P/\rho c^2$ with
present day dark energy density $\OL$, and the present day matter
density is $\OM$.  Each quoted parameter estimate is the median of the
appropriate one-dimensional marginalized posterior 
probability density function (PDF), with the quoted
uncertainties showing, unless otherwise stated, the $16^{\rm th}$ and $84^{\rm
th}$ percentiles (that is, the bounds of a 68\% confidence interval).


\section{Measuring $H_0$ using lensing, stellar dynamics, and lens environment studies}
\label{sec:H0theory}

In this section we briefly review 
the theory of gravitational lensing for $H_0$ measurement
(Section~\ref{sec:H0theory:Lensing}),
describe the
mass-sheet degeneracy (Section \ref{sec:H0theory:MSD}), and present the
dynamics modeling (Section \ref{sec:H0theory:Dynamics}).  Readers
familiar with these subjects can proceed directly to Section
\ref{sec:H0ProbTheory}.


\subsection{Theory of gravitational lensing}
\label{sec:H0theory:Lensing}

For a strong lens system in an otherwise homogeneous Robertson-Walker
universe, the excess time delay of an image at angular position
$\bmath{\theta}=(\theta_1,\theta_2)$ with corresponding source position
$\bmath{\beta}=(\beta_1,\beta_2)$ relative to the case of no lensing is
\be \label{eq:T} 
t(\bmath{\theta},\bmath{\beta}) = \frac {1}{c} \frac{D_{\rm d} D_{\rm s}}{D_{\rm ds}} (1+\zd)\, \phi(\vec{\theta},\vec{\beta}),
\ee
where $\zd$ is the redshift of the lens, $\phi(\vec{\theta},\vec{\beta})$ is
the so-called Fermat potential, and $D_{\rm d}$, $D_{\rm s}$, and $D_{\rm ds}$
are, respectively, the angular diameter distance from us to the lens, from us
to the source, and from the lens to the source. The Fermat potential is
defined as
\be \label{eq:FP}
\phi(\vec{\theta},\vec{\beta})\equiv \left[\frac{(\bmath{\theta}-\bmath{\beta})^2}{2}-\psi(\bmath{\theta}) \right], 
\ee
where the first term comes from the geometric path difference as a result of
the strong lens deflection, and the second term is the gravitational delay
described by the lens potential $\psi(\bmath{\theta})$.  The scaled deflection
angle of a light ray is $\bmath{\alpha}(\bmath{\theta}) = \bmath{\nabla}
\psi(\bmath{\theta})$, and the lens equation that governs the deflection 
of light rays is $\bmath{\beta} = \bmath{\theta} - \bmath{\alpha}(\bmath{\theta})$.

The projected dimensionless surface mass density $\kappa(\bmath{\theta})$ is 
\be \label{eq:psiKappaDiffRelan}
\kappa(\bmath{\theta})=\frac{1}{2}\nabla^2\psi(\bmath{\theta}),
\ee
where 
\be \label{eq:kappa} 
\kappa(\vec{\theta}) = \frac {\Sigma(D_{\rm d} \vec{\theta})} {\Sigma_{\rm cr}} \qquad \mathrm{with} \qquad \Sigma_{\rm cr} = \frac{c^2 D_{\rm s}}{4 \pi G D_{\rm d} D_{\rm ds}},
\ee
and $\Sigma(D_{\rm d} \vec{\theta})$ is the physical projected surface mass density.

The constant coefficient in Equation (\ref{eq:T}) is proportional to
the angular diameter distance and hence inversely proportional to the
Hubble constant.  
We can thus simplify Equation (\ref{eq:T}) to the
following: 
\bea
\label{eq:Tsimp}
t(\bmath{\theta},\bmath{\beta}) & = & \frac{\tdist}{c}\,
\phi(\vec{\theta},\vec{\beta}) \\
\label{eq:TsimpH0}
& \propto & \frac{1}{H_0} \phi(\vec{\theta},\vec{\beta}),
\eea
where $\tdist \equiv (1+\zd) D_{\rm d} D_{\rm s}/D_{\rm ds}$ is referred
to as the time-delay distance. 

Therefore, by modeling the lens potential ($\psi(\bmath{\theta})$)
and the source position ($\bmath{\beta}$), we
can use time-delay lens systems to deduce the value of the Hubble
constant, 
and indeed the other cosmological parameters
that appear in $\tdist$. In this way, strong lensing can be
seen as a kinematic probe of the universal expansion, in the same general
category as SNe Ia and BAO.
Since the principal dependence of $\tdist$ is on $H_0$, we continue
to discuss lenses as a probe of this one parameter; however,
we shall see that the other cosmological parameters play an important role in
the analysis.

Gravitational lens systems with spatially extended source surface brightness 
distributions are of special interest since they provide additional
constraints on the lens potential.  However, in this case, simultaneous
determinations of the source surface brightness and the lens potential are
required.


\subsection{Mass-sheet degeneracy}
\label{sec:H0theory:MSD}

We now 
briefly describe the mass-sheet degeneracy and its relevance to this
research \citep[see e.g.][for details]{FalcoEtal85,SchneiderEtal06}.
As its name suggests, this is a degeneracy in the mass modeling
corresponding to the addition of a  
mass sheet that contributes a convergence and zero shear (and a matching
scaling of the original mass distribution) which leaves the predicted 
image positions unchanged.  A circularly symmetric surface mass
density distribution that is uniform interior to the line of sight is
one example of such a lens.  Suppose we have a lens
model $\kappa_{\rm model}(\vec{\theta})$ that fits the observables of a
lens system (i.e., image positions, flux ratios for point sources, and the
image shapes for extended sources).
A new model described by the
transformation $\kappa_{\rm trans}(\vec{\theta}) = \lambda + (1-\lambda)
\kappa_{\rm model}(\vec{\theta})$, where $\lambda$ is a constant,
would also fit the lensing observables equally well.
The parameter $\lambda$ corresponds physically to the convergence of
the sheet.
Since we might think of including exactly such a parameter to account for
additional physical mass lying along the line of sight, or in the lens plane
to model a nearby group or cluster, it is clear that 
the mass-sheet degeneracy corresponds to a degeneracy between this
external convergence ($\kext$) and the mass normalization of the lens 
galaxy.\footnote{To be specific, the prescription that we adopt for
  combining the effects of many mass sheets at redshifts $z_i$ with
  surface mass densities $\Sigma_i$ is $\kext=\frac{4\pi
    G}{c^2}\displaystyle\sum_i\frac{\Sigma_i(D_i\vec\theta)D_i D_{i{\rm s}}}{D_{\rm s}}$.}

Despite the invariance of the image positions, shapes and relative fluxes
under a mass-sheet transformation, the relative Fermat potential between the
images changes according to $\Delta \phi_{\rm trans}(\vec{\theta},
\vec{\beta}_{\rm trans}) = (1-\lambda) \Delta \phi_{\rm
  model}(\vec{\theta},\vec{\beta}_{\rm model})$.  Therefore, given measured
relative time delays $\Delta t$, which are inversely proportional to $H_0$ and
proportional to the relative Fermat potential (Equation \ref{eq:TsimpH0}), the
transformed model $\kappa_{\rm trans}$ would lead to an $H_0$ that is a factor
$(1-\lambda)$ lower than that of the initial $\kappa_{\rm model}$ (for fixed
$\OM$, $\OL$, and $w$).  In other words, if there is physically any external
convergence $\kext$ due to the lens' local environment or mass structure along the
line of sight
to the lens system that is not incorporated in the lens
modeling, then
\be 
\label{eq:MassSheet:H0bias}
H_0^{\rm{true}}=(1-\kext) H_0^{\rm{model}}.
\ee

This degeneracy is present because lensing observations only deliver
relative positions and fluxes.  The degeneracy can be broken, allowing us to
measure $H_0$, if (i) we
know the magnitude or angular size of the source in absence of lensing,
(ii) we have information on
the mass normalization of the lens,
or (iii) we can compare the measured shear in the lens with the
observed distribution of mass to calibrate $\kext$.
For most of the strong lens systems including \ourlens, 
case (i) does not apply, so
circumventing the mass-sheet degeneracy requires the input of more
information, either about the lensing galaxy itself, or its three-dimensional
environment.
We distinguish between two kinds of mass sheets: \textit{internal}
and \textit{external}.  
Internal mass 
sheets, which are physically associated with the lens galaxy, are due
to nearby, physically associated galaxies, 
groups or clusters which, crucially, 
affect the stellar dynamics of the
lens galaxy. External mass sheets describe mass
distributions that are not physically associated with the
lens galaxy and, by definition, do not affect the stellar dynamics.  
Typically these will 
lie along the line of sight to the lens \citep{FassnachtEtal06}. 
We identify $\kext$ as the net convergence of this external mass sheet.

Two methods for breaking the mass-sheet degeneracy are then: 
\renewcommand{\theenumi}{\roman{enumi}}
\begin{enumerate}
\item \textit{Stellar dynamics of the lens galaxy.}  Stellar dynamics
  can be used jointly with lensing to break the internal mass-sheet degeneracy
  by providing an estimate of the enclosed mass at a radius different
  from the Einstein radius, which is approximately the radius of the
  lensed images from the lens galaxy
  \citep[e.g.,][]{GroginNarayan96a, GroginNarayan96b, TonryFranx99,
    KoopmansTreu02, TreuKoopmans02, BarnabeKoopmans07}.  We note that
  for a given stellar velocity dispersion, there is a degeneracy in
  the mass and the stellar orbit anisotropy (which characterizes the amount of
  tangential velocity dispersion relative to radial dispersion).  Nonetheless,
  the mass-isotropy degeneracy is nearly orthogonal to the mass-sheet
  degeneracy, so a combination of the mass within the effective radius
  (from the stellar velocity dispersion) and the mass within the Einstein
  radius (from lensing) effectively breaks both the mass-isotropy and
  the internal mass-sheet degeneracies. We describe how this works within the
  context of our chosen mass model in Section~\ref{sec:H0theory:Dynamics}
  below.
\item \textit{Studying the environment and the line of sight to the lens galaxy.}  
  Observations of the field around lens galaxies allow a rough picture of the
  projected mass distribution to be built up. Many lens galaxies lie in galaxy
  groups, which can be identified either by their spectra or, more cheaply
  (but less accurately), by their colors and magnitudes.    By modeling the
  mass distribution of the groups and galaxies in the lens plane and along the
  line of sight  to the lens galaxy, one can estimate the external
  convergence~$\kext$  at the redshift of the lens \citep[e.g.][and references
  therein]{MomchevaEtal06, FassnachtEtal06, AugerEtal07}.  The group modeling
  requires (i) identification of the  galaxies that belong to the group of the
  lens galaxy, and (ii) estimates of the group centroid and velocity
  dispersion.  A number of recipes can be followed. For example,
  \citet{KeetonZabludoff04} considered two extremes: (i) the group is
  described by a single smooth mass distribution, and (ii) the masses are
  associated with individual galaxy group members with no common halo.  The
  realistic mass distribution for a galaxy group should be somewhere between
  these two extremes.  
  The experience to date  is that 
  modeling lens environments accurately is very difficult, with uncertainties
  of 100\% typical \citep[e.g.][]{MomchevaEtal06, FassnachtEtal06}.    
  In Section \ref{sec:LensEnv}, we describe an
  alternative approach for quantifying the external convergence in a
  statistical manner: ray-tracing through numerical simulations of large-scale
  structure \citep{HilbertEtal07}.
  In this section 
  we also present a first attempt at tailoring the ray-tracing results 
  to our one line of sight, using the relative galaxy number counts in the
  field.
\end{enumerate}

We emphasize that the mass-sheet degeneracy is simply one of the
several parameter degeneracies in the lens modeling that has been
given a special name.  When power-laws ($\kappa \sim b R^{1-\slope}$,
where $R$ is the radial distance from the lens center, $b$ is the
normalization of the lens, and $\slope$ is the radial slope in the
mass profile) are used to describe the lens mass distribution, one
often finds a $H_0$-$\slope$ degeneracy in addition to the
$H_0$-$b$-$\kext$ (mass-sheet) degeneracy (for fixed $\OM$, $\OL$ and
$w$; more generally, $\tdist$ would be in place of $H_0$).  These two
degeneracies are of course related via $H_0$.  The $H_0$-$\slope$
degeneracy primarily occurs in lens systems with symmetric
configurations due to a lack of information on $\slope$.  In contrast,
lens systems with images spanning a range of radii or with extended
images provide information on $\slope$
\citep[e.g.][]{WucknitzEtal04,DyeEtal08}, and so the $H_0$-$\slope$
degeneracy is broken.  Nonetheless, the $H_0$-$b$-$\kext$ degeneracy
is still present unless we provide information from dynamics and lens
environment studies.  


\subsection{Stellar dynamics modeling}
\label{sec:H0theory:Dynamics}

In order to model the velocity dispersion of the stars in the lens galaxy, we
need a model for the local gravitational potential well in which those stars
are orbiting.  This potential is due to both 
the mass distribution of the lens
galaxy, and also the ``internal mass sheet'' due to 
neighboring groups and galaxies physically associated with the lens, as described in the
previous subsection.  
Recent studies such as the Sloan Lens ACS Survey
(SLACS) and hydrostatic X-ray analyses found that the sum of these internal components can be well-described
by a power law \citep[e.g.][]{TreuEtal06, KoopmansEtal06, 
GavazziEtal07, KoopmansEtal09, HumphreyBuote09}.
With this in mind, we assume that the total (lens plus sheet) mass density distribution is
spherically symmetric and of the form
\be
\label{eq:localdensity}
\rho_{\rm local} = \rho_0 \left(\frac{r_0}{r}\right)^{\slope},
\ee
where $\slope$ is the logarithmic slope of the effective lens density
profile, and $\rho_0 r_0^{\slope}$ is the normalization of the mass
distribution that is determined quite precisely by the lensing, up to
a small offset contributed by the external convergence $\kext$.  This
normalization can be expressed in terms of observable or inferrable
quantities as we show below.

By integrating $\rho_{\rm local}$ 
within a cylinder with radius
given by the Einstein radius $\Rein$, we find 
\bea
M_{\rm local} & = & 4\pi \int_0^\infty dz \int_0^{\Rein} \rho_0 r_0^{\slope} \frac{s\, {\rm d}s}{(s^2 + z^2)^{\slope/2}} \\
\label{eq:Mlocal}
& = & - \rho_0 r_0^{\slope} \frac{\pi^{3/2}
  \Gamma(\frac{\slope-3}{2})
  \Rein^{3-\slope}}{\Gamma(\frac{\slope}{2})}.  
\eea 
However, the mass responsible for creating an Einstein ring is a combination
of this local mass and the external mass contributed along the line of sight, 
so the mass contained within the Einstein ring is
\be
\label{eq:Meq}
M_{\rm Ein} = M_{\rm local} + M_{\rm{ext}} 
\ee 
where $M_{\rm Ein}$ is the mass enclosed within the Einstein radius
$\Rein$ that would be inferred from lensing,\footnote{By definition,
$\Rein$ is the radius within which the total mean convergence is unity.} 
given by 
\be
\label{eq:MEin}
M_{\rm Ein}  =  \pi \Rein^2 \Sigma_{\rm cr}, 
\ee 
and $M_{\rm ext}$ is the mass contribution from $\kext$, 
\be 
\label{eq:Mkext}
M_{\rm{ext}} = \pi \Rein^2 \kext \Sigma_{\rm cr}.  
\ee 
Combining Equations (\ref{eq:Mlocal}), (\ref{eq:Meq}),
(\ref{eq:MEin}), and (\ref{eq:Mkext}), we find
\be
\rho_0 r_0^{\slope} = (\kext - 1)
\Sigma_{\rm cr} \Rein^{\slope-1} \frac{\Gamma(\frac{\slope}{2})}{\pi^{1/2}
  \Gamma(\frac{\slope-3}{2})}.
\ee
Substituting this in Equation (\ref{eq:localdensity}), we obtain
\be
\rho_{\rm local}=(\kext - 1)
\Sigma_{\rm cr} \Rein^{\slope-1} \frac{\Gamma(\frac{\slope}{2})}{\pi^{1/2}
  \Gamma(\frac{\slope-3}{2})} \frac{1}{r^{\slope}}.
\ee
Spherical Jeans modeling can then be employed to infer the
line-of-sight velocity dispersion, $\vdisp^{\rm P}(\slope, \kext,
\beta_{\rm ani}, \OM, \OL, w)$, from $\rho_{\rm local}$
by assuming a model for the
stellar distribution $\rho_*$ \citep[e.g.,][]{BinneyTremaine87}. 
Here, $\beta_{\rm ani}$ is a general anisotropy term that can be expressed
in terms of an anisotropy radius parameters for the stellar velocity
ellipsoid, $\aniso$, in the Osipkov-Merritt formulation
\citep{Osipkov79, Merritt85}:
\be
\beta_{\rm ani} = \frac{r^2}{\aniso^2 + r^2},
\ee
where $\aniso=0$ is pure radial orbits and
$\aniso\rightarrow\infty$ is isotropic with equal radial and
tangential velocity dispersions.
The dependence of $\vdisp^{\rm P}$ 
on $\OM$, $\OL$, and $w$ enters through $\Sigma_{\rm cr}$ and the
physical scale radius of the stellar distribution, but the dependence
on $H_0$ drops out.

We now follow \citet{BinneyTremaine87} to show how the model velocity dispersion is calculated. The three-dimensional radial velocity dispersion $\sigma_{\rm r}$ is found by solving the spherical Jeans equation
\be
\label{eq:Jeansequation}
\frac{1}{\rho_*}\frac{d(\rho_*\sigma_{\rm r})}{dr} + 2\frac{\beta_{\rm ani} \sigma_{\rm r}}{r} = -\frac{GM(r)}{r^2},
\ee
where $M(r)$ is the mass enclosed within a radius $r$ for the total density profile given by Equation (\ref{eq:localdensity}) and with $\rho_*$ given by the Hernquist profile \citep{Hernquist90}
\be
\rho_*(r) = \frac{{\rm{I_0}} a}{2\pi r (r + a)^3},
\ee
where the scale radius $a$ is related to the effective radius $r_{\rm eff}$ by $a = 0.551r_{\rm eff}$ and ${\rm{I_0}}$ is a normalization term. The solution to Equation (\ref{eq:Jeansequation}) is
\bea
\sigma_{\rm r}^2 & = & \frac{4\pi G a^{-\slope} \rho_0 r_0^{\slope}}{3-\slope} \frac{r(r+a)^3}{r^2 + \aniso^2} \cdot \nonumber\\
                 &   & \left( \frac{\aniso^2}{a^2} \frac{{\rm _2F_1}[2+\slope,\slope; 3+\slope; \frac{1}{1+r/a}]}{(2+\slope) (r/a+1)^{2+\slope}} + \right. \nonumber\\
                 &   & \left. \frac{{\rm _2F_1}[3, \slope; 1+\slope; -a/r]}{\slope (r/a)^{\slope}} \right), 
\eea
where ${\rm _2F_1}$ is a hypergeometric function. The model luminosity-weighted velocity dispersion within an aperture $\mathcal{A}$ is then
\be
(\vdisp^{\rm P})^2 = \frac{\int_{\mathcal{A}} [ I_{\rm
    H}(R)\sigma_{\rm s}^2 * \mathcal{P} ]\, R \,{\rm d}R \,{\rm
    d}\theta}{\int_{\mathcal{A}} [ I_{\rm H}(R) * \mathcal{P} ]\, R
  \,{\rm d}R \,{\rm d}\theta},
\ee
where $I_{\rm H}(R)$ is the projected Hernquist distribution \citep{Hernquist90}, both integrands are convolved with the seeing $\mathcal{P}$ as indicated, and the theoretical (that is, before convolution and integration over the spectrograph aperture) luminosity-weighted projected velocity dispersion $\sigma_{\rm s}$ is given by
\be
I_{\rm H}(R)\sigma_{\rm s}^2 = 2\displaystyle \int_R^\infty (1 -
\beta_{\rm ani} \frac{R^2}{r^2})\frac{\rho_*\sigma_{\rm r}^2 r \,{\rm d}r}{\sqrt{r^2 - R^2}}.
\ee
The use of a \citet{Jaffe83} stellar distribution function follows the same derivation.

In the next section, we present the probability theory
for obtaining posterior probability distribution of $H_0$ by combining
the lensing, dynamics and lens environment studies.


\section{Probability Theory}
\label{sec:H0ProbTheory}

We aim to obtain an expression for the posterior probability
distribution of cosmological parameters $H_0$, $\OM$, $\OL$, and $w$
given the various independent data sets of \ourlens.


\subsection{Notations for joint modeling of data sets}
\label{sec:H0ProbTheory:Notation}

We introduce notations for the observed data and the model parameters that
will be used throughout the rest of this paper.  

We have three independent data sets for \ourlens: the time delay
measurements from the radio observations of the four lensed images A,
B, C and D \citep{FassnachtEtal99,
  FassnachtEtal02}, \HST\ Advanced Camera for Surveys (ACS) 
observations associated with
program 10158 (PI:Fassnacht; \citeauthor{SuyuEtal09}~\citeyear{SuyuEtal09}), 
and the stellar velocity dispersion measurement of the primary lens
galaxy G1 (see Section \ref{sec:StellDyn}).  Let $\tdelayVec$ be the
time delay measurements of images A, C and D relative to image B, $\dataVec$
be the data vector of the lensed image surface brightness measurements
of the gravitational lensed image, and $\vdisp$ be the stellar
velocity dispersion measurement of the lens galaxy.

As shown in Section~\ref{sec:H0theory:Lensing},
information on $H_0$, $\OM$, $\OL$, and $w$ comes primarily from the relative
time delays between the images, which is a product of the 
time-delay distance $\tdist$  
and the Fermat potential difference.  The Fermat potential is
determined by the lens potential and the source position that is given by the
lens equation. Therefore, the first step is to model the lens system using the
observed lensed image $\dataVec$.  In order to model the lens mass
distribution using the extended source information, we need to model the
point-spread function (PSF) $\blurSet$, image covariance matrix $\imCM$, lens
galaxy light $\glightVec$, and dust $\dustSet$ (if present)
\citep[e.g.][]{SuyuEtal09}.  We collectively denote these discrete 
models associated
with the lensed image processing as
$\imModelSet=\{\blurSet,\imCM,\glightVec,\dustSet\}$.  
We explored a representative subspace of models
$\imModelSet$ in \paperI, using the Bayesian evidence from the ACS data
analysis to quantify the appropriateness of each model tested.
Given a particular image processing model,
we can infer the parameters of the lens
potential and the source surface brightness distribution 
from the ACS data $\dataVec$.
The data models are denoted by $\imModelNum_j =
\imModelNum_2,\ldots,\imModelNum_{11}$  for Models 2--11 in
\paperI.

The lens potential can be simply parametrized by, for example, a
singular power-law ellipsoid (SPLE) with surface mass density
\be
\label{eq:sple}
\kappa(\theta_1,\theta_2) = b \left[\theta_1^2+\frac{\theta_2^2}{q^2} \right]^{(1-\slope)/2},
\ee
where $q$ is the axis ratio, $b$ is the lens strength that determines
the Einstein radius ($R_{\rm Ein}$), and $\slope$ is the radial slope
\citep[e.g.][]{Barkana98,KoopmansEtal03}.
The distribution is then translated (with two parameters for the
centroid position) and rotated by the position angle parameter.  
There is no need to include an 
external convergence parameter in the mass
distribution during the lens modeling
since we cannot determine it due to the mass-sheet
degeneracy.
Instead, we explicitly 
incorporate the external convergence in
the Fermat potential later on, taking into account the interplay among this
parameter, the slope, and the normalization of the effective lens mass
distribution.  We collectively label all the
parameters of the simply-parametrized model by $\lenspars$, except
for the radial slope $\slope$. 

Alternatively, the
lens potential can be described on a grid of pixels, especially when
the source galaxy is spatially extended (which provides additional
constraints on the lens potential).  We focus on this case; in
particular, we decompose the lens potential into an initial
simply-parametrized SPLE model $\psiI(\slope,\lenspars)$ and
grid-based potential corrections denoted by the vector $\dpsiVec$.  The
final potential, which is on the same grid of pixels as the
corrections, is $\psiVec = \psiIVec(\slope,\lenspars) + \dpsiVec$,
where $\psiIVec(\slope,\lenspars)$ is the vector of initial potential
values evaluated at the grid points.  Furthermore, we also describe the
extended source surface brightness distribution on a (different) grid of pixels by the
vector $\srVec$.  The determination of the source surface brightness
distribution given the lens potential model is a regularized linear
inversion.  The strength and form of the regularization are denoted by
$\lambda$ and $\regSet$, respectively.  The procedure for obtaining
the pixelated potential corrections and the corresponding extended
source surface brightness distribution is iterative and is described in detail
in \paperI.  We highlight that the resulting (iterated) pixelated lens
potential model is not limited by the parametrization of the initial
SPLE model -- tests of this method in \paperI\ showed that when the
iterative procedure converged, the true potential was reconstructed
irrespective of the initial model.

The resulting lens potential allows us to compute the Fermat potential
$\fp$ at each image position, up to a factor of $(1-\kext)$.
Combining the Fermat potential with a value of $\tdist$ computed given the 
cosmological parameters \{$H_0$, 
$\OM$, $\OL$, $w$\} provides us with 
predicted values of the image time delays, $\tdelay^{\rm P}$. 

The dynamics modeling of the galaxy is performed following 
Section \ref{sec:H0theory:Dynamics}.  
By construction, the power-law profile for the dynamics modeling with
slope $\slope$ matches the radial profile of the SPLE.  Although
spherical symmetry is assumed for the dynamics modeling, a suitably
defined Einstein radius from the lens modeling leads to $R_{\rm Ein}$
and $M_{\rm Ein}$ that are independent of $q$ and are directly
applicable to the spherical dynamics modeling
\citep[e.g.][]{KoopmansEtal06}.  Furthermore, the results from SLACS
based on spherical dynamics modeling \citep{KoopmansEtal09} agree with
those from a more sophisticated two-dimensional kinematics analyses of
six SLACS lenses \citep{CzoskeEtal08, BarnabeEtal09}, indicating that
spherical dynamics modeling for B1608+656 is sufficient.  The
predicted velocity dispersion is dependent on six parameters: 
1) the effective lens mass distribution profile slope $\slope$, 
2) the external convergence $\kext$, 
3) the anisotropy radius $\aniso$, and then the cosmological parameters 4)
$\OM$, 5) $\OL$, and 6) $w$.  

By combining lensing, dynamics, and lens environment studies, we can break
the $\tdist$-$\kext$ degeneracy to obtain a probability distribution for the 
cosmological parameters \{$H_0$, $\OM$, $\OL$, $w$\} 
given the data sets.  In the inference, we assume that the redshifts of the
lens and source galaxies are known exactly for the computation of $\tdist$.
This is approximately true for \ourlens, which has spectroscopic measurements
for the redshifts \citep{MyersEtal95, FassnachtEtal96} --- an
uncertainty of $0.0003$ on the redshifts translates to $<0.2\%$
in time-delay distance, and hence $H_0$ for fixed $\OM$, $\OL$ and $w$.  
By imposing sensible
priors on \{$H_0$, $\OM$, $\OL$, $w$\} from other independent experiments such as
WMAP5, we can marginalize the distribution to obtain the 
posterior probability distribution for $H_0$.


\subsection{Constraining cosmological parameters}
\label{sec:H0ProbTheory:CosParamConstraint}

In this section, we describe the probability theory for inferring
cosmological parameters from the  \ourlens\ data sets.
Readable introductions to this type of analysis can be found in the books by
\citet{sivia} and \citet{mackay}; we use notation consistent
with that in \paperI.

Our goal is to obtain the posterior PDF
for the model parameters $\pars$ given the three independent data sets
\{$\tdelayVec$, $\dataVec$, $\vdisp$\}:
\be
\label{eq:ParsPosterior:sec}
 P(\pars|\tdelayVec,\dataVec,\vdisp) \propto P(\tdelayVec|\pars)P(\dataVec|\pars)P(\vdisp|\pars)P(\pars),
\ee
where the parameters $\pars$ consist of all the model parameters
for obtaining the predicted data sets described in Section
\ref{sec:H0ProbTheory:Notation}: $\slope$, $\kext$, $\lenspars$, $\dpsiVec$,
$\srVec$, $\imModelSet$, $\aniso$, $H_0$, $\OM$,
$\OL$, $w$.  For notational simplicity, we denote the cosmological parameters as
$\cosmopars=\{H_0, \OM, \OL, w\}$. In Equation
(\ref{eq:ParsPosterior:sec}), the dependence on $\zs$ and $\zd$ are implicit.

To obtain the PDF of cosmological parameters $\cosmopars$, we
marginalize Equation (\ref{eq:ParsPosterior:sec}) over all parameters
apart from $\cosmopars$:
\bea
\label{eq:CosmoparsPosterior:sec}
 P(\cosmopars|\tdelayVec,\dataVec,\vdisp) &\propto& \int {\rm d}\slope\ {\rm d}\kext\ {\rm d}\lenspars\ {\rm d}\dpsiVec\ {\rm d}\srVec\ {\rm d}\imModelSet\ {\rm d}\aniso \cdot \nonumber\\
& & \overbrace{P(\tdelayVec|\pars)P(\dataVec|\pars)P(\vdisp|\pars)}^{\rm likelihood} \cdot \nonumber\\
& & \overbrace{P(\cosmopars,\slope,\kext,\lenspars,\dpsiVec,\srVec ,\imModelSet,\aniso)}^{\rm prior}.
\eea
In the following subsection, we discuss each of the three terms in the joint
likelihood function in turn.


\subsection{Likelihoods}
\label{sec:H0ProbTheory:Like}

Each of the three
likelihoods in Equation (\ref{eq:CosmoparsPosterior:sec}) generally
depends only on a subset of the parameters $\pars$.  Specifically,
dropping independences, we have
$P(\tdelayVec|\pars)=P(\tdelayVec|\cosmopars,\slope,\kext,\lenspars,\dpsiVec,\imModelSet)$,
$P(\dataVec|\pars)=P(\dataVec|\slope,\lenspars,\dpsiVec,\srVec,\imModelSet)$,
and $P(\vdisp|\pars)=P(\vdisp|\OM,\OL,w,\slope,\kext,\aniso)$. 

For \ourlens, we can simplify and drop independences further in the time
delay likelihood $P(\tdelayVec|\pars)$ by expressing the relative
Fermat potential (relative to image B for the images A, C 
and D) as
\be
\label{eq:fprelation:sec}
\fpdiff(\slope, \kext, \imModelSet) = (1-\kext) q(\slope,\imModelSet),
\ee
and writing the $i^{\rm th}$ (AB, CB or DB) predicted time delay as 
\be
\label{eq:tddef:sec}
  \tdelay_i^{\rm P} = \frac{1}{c}\tdist(\zd,\zs,\cosmopars) \cdot \fpdiff_i(\slope,\kext, \imModelSet)
\ee 
(see Appendix~\ref{app:H0ProbTheory} for details).  
The resulting likelihood is 
\bea
\label{eq:tdelayLikeSimp:sec}
\lefteqn{ P(\tdelayVec | \zd,\zs, \cosmopars, \slope,\kext,\imModelSet)} \nonumber\\
& & = \prod_{i=1}^{3} P(\tdelay_i | \zd,\zs, \cosmopars, \slope,\kext, \imModelSet),
\eea
where we assume that the three time delay measurements are independent, and
that each
$P(\tdelay_i | \zd,\zs, \cosmopars, \slope,\kext, \imModelSet)$
is given by the PDF in \citet{FassnachtEtal02}.

The pixelated lens potential and source surface brightness 
reconstruction allows us to compute
\bea
\label{eq:SrEvidDef:sec}
P(\dataVec | \slope, \lenspars, \dpsiMPVec, \imModelSet) &= \int &{\rm d}\srVec\ P(\dataVec | \slope, \lenspars, \dpsiMPVec, \srVec, \imModelSet) \cdot \nonumber\\
& & P(\srVec | \lambda, \regSet),
\eea
by marginalizing out the source surface brightness $\srVec$.  The
most probable potential correction, $\dpsiMPVec$, is the result of the
pixelated potential reconstruction method.  
The likelihood for the lens parameters, 
$P(\dataVec | \slope, \lenspars, \dpsiMPVec, \imModelSet)$, is also
the Bayesian evidence of the source surface brightness reconstruction;
the analytic expression for this likelihood is given by Equation (19)
in \citet{SuyuEtal06}.  Part of the 
marginalization in Equation (\ref{eq:CosmoparsPosterior:sec}) can be
simplified via
\bea
 & & \int {\rm d}\dpsiVec\ {\rm d}\srVec\ {\rm d}\imModelSet\ P(\dataVec | \slope, \lenspars, \dpsiVec, \srVec, \imModelSet)\cdot\nonumber\\
& & \ \ \ \ \ \  P(\srVec | \lambda, \regSet) P(\imModelSet) P(\dpsiVec)\nonumber \\
\label{eq:CosmoparsPosteriorSimp1:sec}
& \propto  & \sim P(\dataVec | \slope, \lenspars, \imModelSet=\imModelFive),
\eea
under various assumptions stated in
Appendix~\ref{app:H0ProbTheory} that are either justified
in Paper I or will be shown to be valid in Section \ref{sec:LensModel:Result}.
In essence, we find that the ACS data models that give acceptable fits are all
equally probable within their errors, making conditioning on
$\imModelFive$ (i.e., setting $\imModelSet=\imModelFive$, where
$\imModelFive$ is Model 5 in \paperI\ for the lensed image processing)
approximately equivalent to marginalizing over all models $\imModelSet$.

Furthermore, we can 
marginalize out the parameters of the smooth lens model~$\lenspars$ 
separately:
\bea
\label{eq:MargSlopePosterior:sec}
P(\slope | \dataVec, \imModelSet=\imModelFive) &\propto& \int {\rm d}\lenspars\ P(\dataVec | \slope, \lenspars, \imModelSet=\imModelFive)\cdot\nonumber\\
& & \ \ P_{\rm no\, ACS}(\slope)\ P(\lenspars).
\eea
(See Appendix~\ref{app:H0ProbTheory} for details of the assumptions involved.)
We see that the resulting PDF,
$P(\slope | \dataVec, \imModelSet=\imModelFive)$, can itself be treated as
a prior on the slope $\slope$. Without the ACS data $\dataVec$, this
distribution will default to the lower level prior~$P_{\rm no\,
  ACS}(\slope)$.  For the rest of
this section we refer only to the generic prior~$P(\slope)$, keeping in mind
that this distribution may or may not include the information from the ACS
data. This will allow us to isolate the influence of the ACS
data on the final results, when we compare the PDF in
Equation (\ref{eq:MargSlopePosterior:sec})
with some alternative choices of $P(\slope)$.

For the velocity dispersion likelihood, the predicted velocity dispersion 
$\vdisp^{\rm P}$ as a function of the parameters described in
Section \ref{sec:H0ProbTheory:Notation} is
\be
\vdisp^{\rm P} = \vdisp^{\rm
  P}(\OM,\OL,w,\slope,\kext,\aniso|\zd,\zs, r_{\rm eff},R_{\rm Ein}),
\ee 
where the effective radius, $r_{\rm eff}$, the
Einstein radius, $R_{\rm Ein}$, and the mass enclosed within the
Einstein radius, $M_{\rm Ein}$, are fixed.  The effective radius is
fixed by observations, and $R_{\rm Ein}$ and $M_{\rm Ein}$ are the
quantities that lensing delivers robustly. The uncertainty in the dynamics
modeling due to the error associated with $r_{\rm eff}$, $R_{\rm Ein}$
and $M_{\rm Ein}$ is negligible compared to the uncertainties
associated with $\kext$.  The likelihood
function for $\vdisp$ is a Gaussian:
\bea
\label{eq:vdispLikelihood:sec}
\lefteqn{P(\vdisp | \OM,\OL,w,\slope,\kext,\aniso)}  \nonumber \\ 
& & = \frac{1}{\sqrt{2\pi\sigma_{\vdisp}^2}} \exp{\left[-\frac{(\vdisp -
\vdisp^{\rm P})^2}{2\sigma_{\vdisp}^2}\right]}.
\eea

Finally then, we have the following simplified version of Equation (\ref{eq:CosmoparsPosterior:sec}),
where the posterior PDF has been successfully compartmentalized into
manageable pieces:
\bea
\label{eq:CosmoparsPosteriorFullSimp2:sec}
P(\cosmopars|\tdelayVec,\dataVec,\vdisp) & \propto & \int {\rm d}\slope\, {\rm d}\kext\, {\rm d}\aniso\cdot\nonumber\\
& & \ \ \ \  P(\tdelayVec | \zd, \zs, \cosmopars, \slope,\kext,\imModelSet=\imModelFive)\cdot\nonumber\\
& & \ \ \ \ P(\vdisp | \OM,\OL,w,\slope,\kext,\aniso)\cdot\nonumber\\
& & \ \ \ \  P(\slope)\, P(\kext)\, P(\aniso)\, P(\cosmopars).
\eea

Sections 4 to 7 address the specific forms of the likelihoods and the
priors in Equation (\ref{eq:CosmoparsPosteriorFullSimp2:sec}).  In
particular, in the next section, we focus on the lens modeling of
\ourlens\ which will justify the assumptions mentioned above
and provide both the time delay likelihood
and the ACS $P(\slope)$ prior.


\section{Lens model of \ourlens}
\label{sec:LensModel}

\begin{figure}
\begin{center}
\includegraphics[width=75mm]{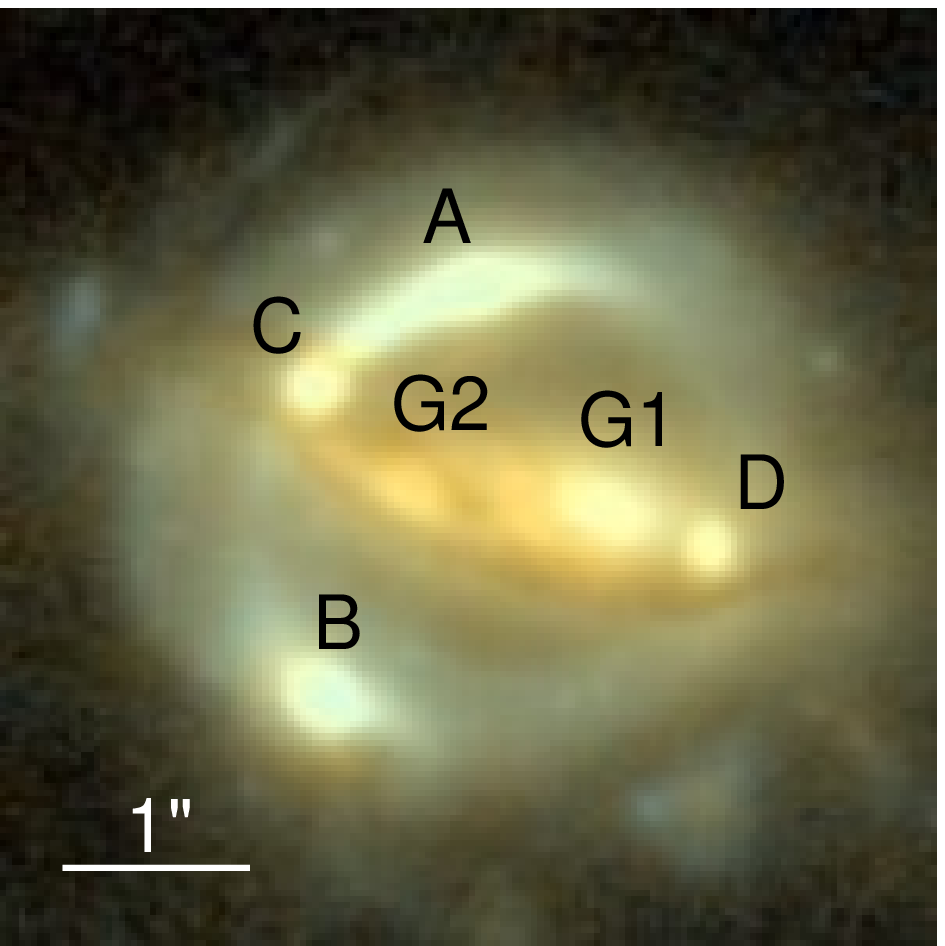}
\caption[\HST\ ACS image of \ourlens]{\label{fig:B1608color}
\HST\ ACS image of \ourlens\ from 11 orbits in F814W and 9 orbits in
F606W.  North is up and east is left.  The lensed images of the source galaxy are labeled by A,
B, C, and D, and the two lens galaxies are G1 and G2. 1 arcsec corresponds to
approximately 7~kpc at the redshift of the lens.}
\end{center}
\end{figure}

The quadruple-image gravitational lens \ourlens\ was discovered in the
Cosmic Lens All-Sky Survey (CLASS) \citep{MyersEtal95, BrowneEtal03, 
MyersEtal03}.
Figure \ref{fig:B1608color} is an image of \ourlens, showing the
spatially extended source surface brightness 
distribution (with lensed images
labeled by A, B, C, and D) and two interacting galaxy lenses (labeled
by G1 and G2).  The redshifts of
the source and the lens galaxies are, respectively, $z_{\rm s}= 1.394$
\citep{FassnachtEtal96} and $z_{\rm d}= 0.6304$
\citep{MyersEtal95}.\footnote{We assume that the redshift of G2 is the
same as G1.}  We note that the lens galaxies are in a group with all
galaxy members in the group lie within $\pm300\kms$ of the mean
redshift \citep{FassnachtEtal06}.  Thus, even a conservative
limit of $300\kms$ for the peculiar velocity of \ourlens\ relative to
the Hubble flow would only change $\tdist$ by $0.5\%$.
As we will see, this is not significant compared to the
systematic error associated with $\kext$.
This system is special in that
the three relative time delays between the four images were measured
accurately with errors of only a few percent: $\Delta t_{\rm
  AB}=31.5^{+2.0}_{-1.0} \rm{\ days}$, $\Delta t_{\rm CB}= 36.0^{+1.5}_{-1.5} 
  \rm{\ days}$, and $\Delta t_{\rm DB}= 77.0^{+2.0}_{-1.0} \rm{\ days}$
\citep{FassnachtEtal99, FassnachtEtal02}.  The additional constraints
on the lens potential from the extended source analysis 
and the accurately measured time delays between
the images make \ourlens\ a good candidate to measure $H_0$ with
few-percent precision.  However, the presence of dust and interacting
galaxy lenses (visible in Figure \ref{fig:B1608color}) complicate this
system.  In \paperI, we presented a comprehensive analysis
that took into account the extended source surface
brightness distribution, interacting galaxy lenses, and the presence of
dust for reconstructing the lens potential.  In the following
subsections, we summarize the data analysis and lens modeling from
\paperI, and present the resulting Bayesian evidence values (needed in
Equation (\ref{eq:MargSlopePosterior:sec})) from the lens modeling.

\subsection{Summary of observations, data analysis, and lens modeling in Paper
I}
\label{sec:LensModel:ObsAnal}

Deep \HST\ ACS observations on \ourlens\ in F606W and F814W filters 
were taken specifically to
obtain high signal-to-noise ratio images of the lensed source emission.

In Paper I, we investigated a representative sample of PSF, dust, and lens
galaxy light models in order to extract the Einstein ring for the lens
modeling.  Table~\ref{tab:EvidSPLE1D} lists the various PSF and dust
models, and we refer the readers to Paper I for details of each model.

The resulting dust-corrected, galaxy-subtracted F814W image 
allowed us to model both the lens potential and source surface brightness 
on grids of pixels based on an iterative and perturbative
potential reconstruction scheme.  This method requires an initial
guess potential model that would ideally be close to the true model.
In Paper I, we adopt the SPLE1+D (isotropic) model from
\citet{KoopmansEtal03} as the initial model, which is the most
up-to-date, simply-parametrized model combining both lensing and stellar
dynamics.  In the current paper, we
additionally investigate the dependence on the initial model by
describing the lens galaxies as SPLE models for a range of slopes
($\slope=1.5, 1.6, \ldots, 2.5$).  Contrary to the SPLE1+D (isotropic)
model, the parameters for the SPLE models with variable slopes
are constrained by lensing data only, without the velocity dispersion
measurement.  

The source reconstruction
provides a value for the Bayesian evidence, $P(\dataVec | \slope, \lenspars,
\dpsiVec, \imModelSet)$, which can be used for model comparison (where
model refers to the PSF, dust, lens galaxy light, and lens potential
model).  The reconstructed lens potential (after the
pixelated corrections $\dpsiVec$) for each data model
(PSF, dust, lens galaxy light) leads to three estimates of the Fermat
potential differences between the image positions.  These are
presented in the next subsection for the representative set of PSF, dust,
lens galaxy light, and pixelated potential model.


\subsection{Lens modeling results}
\label{sec:LensModel:Result}

In \paperI, we successfully used a pixelated
reconstruction method to model small deviations from a smooth 
lens potential model of \ourlens.  
The resulting source surface brightness distribution is well-localized, and the
most probable 
potential correction~$\dpsiMPVec$ has angular structure approximately following a $\cos \phi$
mode with amplitude $\sim2\%$. The $\cos 2\phi$ mode, which could mimic an
additional external shear or lens mass distribution 
ellipticity, has a lower amplitude still, indicating that the
smooth model of \citet{KoopmansEtal03} --- which includes an external shear of
$\simeq 0.08$ --- is giving an adequate account of the
extended image light distribution. This was the main result of \paperI.
The key ingredient in the ACS prior for the lens density profile slope
parameter~$\slope$ (Equation (\ref{eq:MargSlopePosterior:sec})) coming from this
analysis is the
likelihood $P(\dataVec | \slope, \imModelSet)$. For a particular choice of
slope~$\slope$ and data model~$\imModelSet$,
this is just the evidence value resulting from the \paperI\ reconstruction.
In this section, our objective is to use the results of this analysis to
obtain $P(\slope | \dataVec)$ and $\fpdiff(\slope, \kext)$, 
marginalizing over a 
representative sample of data models.


\subsubsection{Marginalization of the data model}

Table~\ref{tab:EvidSPLE1D} shows the
results of the pixelated potential reconstruction at fixed density 
slope in the initial smooth 
lens potential model, for various data models $\imModelSet$.
Specifically, we used the SPLE1+D (isotropic) model in
\citet{KoopmansEtal03} with $\slope=2.05$.
The uncertainties in the log evidence in Table \ref{tab:EvidSPLE1D} were
estimated as  $\sim0.03\times10^4$ for the log evidence values before
potential correction, and $\sim0.05\times10^4$ for the log evidence values
after potential correction.  

We see a clear division between models with high and low evidence values,  the
two groups being separated by a very large factor in probability.
Assuming that all the data models $\imModelSet$ are equally probable {\it a
priori},  the contribution to the marginalized distribution
$P(\cosmopars|\tdelayVec,\dataVec,\vdisp)$ 
(Equation (\ref{eq:CosmoparsPosterior:sec})) from these lower-evidence
models will be negligible.

The physical difference between these evidence-ranked data models is
in the dust correction: the 2-band dust models are found to be less
probable than the 3-band dust models.  It is useful to quantify the
systematic error that would occur with the use of 2-band dust models
(which was avoided from the evidence ranking) in terms of the $H_0$
value implied by the system. For this simple error estimation we use
Equation (\ref{eq:Tsimp}) and assert $\OM=0.3$, $\OL=0.7$, $w=-1$
and zero external convergence,
as a fiducial reference cosmology \citep{KoopmansEtal03}.
The implied Hubble
constants are shown in the final four columns of Table~\ref{tab:EvidSPLE1D}.
We see that the disfavored use of the 2-band dust maps would have led to 
values of $H_0$ some 15\% lower than that inferred from the 3-band maps.

We note that the evidence values of each of the
3-band dust map models $\imModelSet$ are the same within their uncertainties.
We can also see that for good data models, specifically
$\imModelSet=\imModelFive$,  the three $H_0$ values have low scatter: these
lens models are  internally self-consistent. Furthermore, the scatter between
the values for the different good data models is also low: the high evidence
data models consistently return the same Hubble constant. This is the basis
for the approximations (in Section \ref{sec:H0ProbTheory:Like}  and Appendix
\ref{app:H0ProbTheory}) that the likelihood
$P(\tdelayVec|\pars)$ is effectively constant with the 3-band
dust map models $\imModelSet$.  Assuming
that we have indeed obtained the optimal set of $\imModelSet$, we can
approximate the likelihoods in Equations (\ref{eq:MargSlopePosterior:sec})
and (\ref{eq:CosmoparsPosteriorFullSimp2:sec}) as being evaluated for model
$\imModelFive$. 


\subsubsection{Effects of the potential corrections}

Having approximately marginalized out $\imModelSet$ by conditioning on 
$\imModelFive$,   we now consider the impact of the potential corrections
discussed in \paperI. In particular, we seek the likelihood for the density
profile slope parameter~$\slope$, $P(\dataVec | \slope=\slope_i, \lenspars,
\dpsiMPVec,\imModelSet=\imModelFive)$.  We characterize this function on a
grid of slope values in the range of $\slope=1.5,1.6\ldots,2.5$, first
re-optimizing the parameters of the smooth lens model, and then computing the
source reconstruction evidences both with and without potential correction.
These are
tabulated in Table~\ref{tab:slopeEvid}. We again compute the Fermat
potential differences and implied Hubble constant values as before.

The spread of the three implied $H_0$ values  at fixed density slope is again
small: we conclude that  the internal self-consistency of the lens model
depends on the data model but not~$\slope$. The table also shows that the smooth
SPLE model provides a good estimate of the relative Fermat potentials. Indeed,
this was the principal conclusion of \paperI. The relative thickness of the
arcs is sensitive to the SPLE density profile slope~$\slope$, as can be seen 
in the first two columns of Table~\ref{tab:slopeEvid}: the evidence clearly
favors $\slope \simeq 2.05$, as previously found by \citet{KoopmansEtal03}.
Indeed, exponentiating this gives quite a sharply peaked function, which we
return to below.

How is the potential correction then affecting the model? In
Table~\ref{tab:slopeEvid} we can see that the corrected potential leads to
nearly the same evidence value ($P(\dataVec | \slope=\slope_i, \lenspars,
\dpsiMPVec,\imModelSet=\imModelFive)$) for a wide range of underlying density
slopes, and yet barely changes the relative Fermat potential values. 
The unchanging nature of the Fermat potential is due to the curvature type of
regularization on the potential corrections suppressing the addition of
mass within the potential reconstruction annulus.  
From \citet{Kochanek02},
the relative Fermat potential depends only on the mean surface mass density
enclosed in the annulus between the images, to first order in
$\delta R/\langle R\rangle$, where $\delta R$ is the difference in the radial
distance of the image locations from the effective center of the lens galaxies
and $\langle R\rangle$ is the mean radius of
the images.  The mean surface mass density depends on the slope of the initial
SPLE model (hence the trend we see in relative Fermat potential in the
left-hand side of Table~\ref{tab:slopeEvid}), but not on the potential
corrections due to the curvature regularization imposed.  Therefore, to
first order in $\delta R/\langle R\rangle$, the Fermat potential depends
only indirectly on $\slope$ via the mean surface mass density. 
The second order term is very small
--- it has a prefactor of 1/12 and for \ourlens, $(\delta R/\langle R\rangle)^2
\sim 0.1$.  Therefore, for good and self-consistent data models, the potential
corrections $\dpsiMPVec$ do not change the Fermat potential significantly.  

The right-hand side of Table~\ref{tab:slopeEvid}, 
where a wide range of initial slope values
provide good fits to the data, is therefore 
effectively a manifestation of the
mass-sheet degeneracy.  One can understand the effect of the potential
corrections as making {\it local corrections to the effective density profile 
slope in order
to fit the ACS data}.
The change in slope by the pixelated
corrections would create a deficit/surplus of mass in the annulus,
which the pixelated potential corrections then add/subtract back into
the annulus in the form of a constant mass sheet to (i) enforce the
prior (no net addition of mass within annulus) and (ii) continue to
fit the arcs equally well.  

We conclude that the value of the potential correction analysis is in
demonstrating that the double SPLE model for \ourlens\ is, despite the
system's complexity, a good model for the high fidelity \HST\ data.
The corrections are small in magnitude ($\simeq 2\%$ relative to the
initial SPLE model), and the inclusion of
the $\dpsiVec$ neither
significantly reduces the dispersion in implied $H_0$ values between the image
pairs, nor alters the rank order of the data models.
We therefore
use the information on the slope of the initial SPLE model
from the ACS data {\it without potential corrections,}
thus using the information on the relative thickness of the lensed
extended images clearly present.   
How we derive our estimate for $P(\dataVec | \slope, \imModelSet)$ 
from
column 2 of Table~\ref{tab:slopeEvid} is described next.

\begin{table*}
\begin{center}
\caption{\label{tab:EvidSPLE1D} log evidence values and relative Fermat potential values before and after the pixelated potential reconstruction for various data models with the SPLE1+D (isotropic) initial model}
\begin{tabular}{|c|c|c|c|c|c|c|c|c|c|c|c|c|c|}
\hline
\multicolumn{3}{|c|}{Data Model} & \multicolumn{1}{|c|}{Initial Potential} & \multicolumn{8}{|c|}{Corrected Potential} \\
\hline
Model & PSF & dust & log P   & log P & $\fpdiff^{\rm{AB}}$ & $\fpdiff^{\rm{CB}}$  & $\fpdiff^{\rm{DB}}$ & $H_0^{\rm{AB}}$  & $H_0^{\rm{CB}}$  & $H_0^{\rm{DB}}$ & $\bar{H_0}$ \\
      &     &      & $(\times10^4)$ & $(\times10^4)$&  &  &  &  &  &  &  \\
\hline        
5 & B1 & 3-band    & $1.56$ & $1.77$ & 0.244 & 0.279 & 0.575 & 78.1 & 78.1 & 75.1 & $77.1 \pm 1.7$ \\
9 & C  & B1/3-band & $1.56$ & $1.76$ & 0.240 & 0.280 & 0.563 & 76.7 & 78.3 & 73.5 & $76.2 \pm 2.4$ \\
3 & C  & 3-band    & $1.60$ & $1.76$ & 0.243 & 0.277 & 0.570 & 77.6 & 77.5 & 74.4 & $76.5 \pm 1.8$ \\
2 & drz& 3-band    & $1.48$ & $1.75$ & 0.238 & 0.278 & 0.548 & 76.0 & 77.7 & 71.6 & $75.1 \pm 3.1$ \\
7 & B2 & 3-band    & $1.55$ & $1.75$ & 0.237 & 0.274 & 0.571 & 75.7 & 76.7 & 74.6 & $75.7 \pm 1.0$ \\ 
\hline
11& B1 & no dust   & $1.27$ & $1.72$ & 0.229 & 0.263 & 0.576 & 73.2 & 73.6 & 75.3 & $74.0 \pm 1.1$ \\
10& B1 & C/2-band  & $1.36$ & $1.61$ & 0.193 & 0.227 & 0.565 & 61.8 & 63.5 & 73.8 & $66.4 \pm 6.4$ \\ 
4 & C  & 2-band    & $1.40$ & $1.58$ & 0.199 & 0.234 & 0.560 & 63.6 & 65.6 & 73.1 & $67.4 \pm 5.0$ \\ 
6 & B1 & 2-band    & $1.10$ & $1.41$ & 0.196 & 0.226 & 0.559 & 62.5 & 63.2 & 73.0 & $66.2 \pm 5.8$ \\ 
8 & B2 & 2-band    & $1.23$ & $1.40$ & 0.201 & 0.234 & 0.556 & 64.3 & 65.4 & 72.7 & $67.4 \pm 4.5$ \\ 
\hline
\hline
 & & & &  \multicolumn{8}{|c|}{$\fpdiff$ and $H_0$ values from initial SPLE1+D (isotropic)} \\
\hline
 & & & & & 0.243 & 0.271 & 0.575 & 77.7 & 75.8 & 75.1 & $76.2 \pm 1.3$ \\
\hline
\end{tabular}
\end{center}
Notes --- The uncertainties in the log evidence before and after the potential
corrections are $\sim0.03\times10^4$ and $\sim0.05\times10^4$, respectively.
The relative Fermat potentials are in units of ${\rm arcsec}^2$, and the 
$H_0$ values are in units of $\kmsMpc$.  
The $\bar{H_0}$ values are the
mean and standard deviation from the mean of the three estimates obtained
using the initial/corrected potential and the three time delays, without 
taking into account the uncertainties associated with the time delays. 
These $H_0$ values assume $\OM=0.3$, $\OL=0.7$ and $w=-1$, and are listed
purely to aid the digestion of the $\fpdiff$ values.
The full analysis for obtaining the probability distribution for the
cosmological parameters is described in Section \ref{sec:H0}.
\end{table*}

\begin{table*}
\begin{center}
\caption{\label{tab:slopeEvid} log evidence value before and after the pixelated potential reconstruction for initial models with various slope using PSF-B1 and the 3-band dust map ($\imModelSet$=Model 5)}
\begin{tabular}{|c|c|c|c|c|c|c|c|c|c|c|c|c|c|c|c|c|}
\hline
 & \multicolumn{8}{|c|}{Initial Potential} & \multicolumn{8}{|c|}{Corrected Potential} \\
\hline
$\slope$ & log P & $\fpdiff^{\rm{AB}}$ & $\fpdiff^{\rm{CB}}$  & $\fpdiff^{\rm{DB}}$ & $H_0^{\rm{AB}}$  & $H_0^{\rm{CB}}$  & $H_0^{\rm{DB}}$  & $\bar{H_0}$  & log P & $\fpdiff^{\rm{AB}}$ & $\fpdiff^{\rm{CB}}$  & $\fpdiff^{\rm{DB}}$ & $H_0^{\rm{AB}}$  & $H_0^{\rm{CB}}$  & $H_0^{\rm{DB}}$  & $\bar{H_0}$  \\
 & ($\times10^4$) & & & & & & & & ($\times10^4$) & & & & & & & \\
\hline
1.5 & 1.38 & 0.125 & 0.139 & 0.287 & 40.2 & 39.0 & 37.6 & $38.9 \pm 1.3$ & 1.73 & 0.130 & 0.143 & 0.290 & 41.7 & 40.2 & 38.0 & $39.9 \pm 1.9$\\
1.6 & 1.48 & 0.147 & 0.163 & 0.338 & 47.2 & 45.8 & 44.3 & $45.7 \pm 1.4$ & 1.77 & 0.150 & 0.170 & 0.349 & 48.1 & 47.6 & 45.6 & $47.1 \pm 1.3$\\
1.7 & 1.52 & 0.174 & 0.193 & 0.403 & 55.5 & 54.0 & 52.7 & $54.0 \pm 1.4$ & 1.75 & 0.178 & 0.201 & 0.417 & 57.0 & 56.2 & 54.5 & $55.9 \pm 1.3$\\
1.8 & 1.54 & 0.190 & 0.211 & 0.442 & 60.8 & 59.1 & 57.7 & $59.2 \pm 1.5$ & 1.77 & 0.194 & 0.215 & 0.457 & 61.9 & 60.2 & 59.7 & $60.7 \pm 1.2$\\
1.9 & 1.58 & 0.210 & 0.234 & 0.491 & 67.1 & 65.4 & 64.1 & $65.6 \pm 1.4$ & 1.76 & 0.210 & 0.237 & 0.510 & 67.3 & 66.4 & 66.6 & $66.8 \pm 0.5$\\
2.0 & 1.60 & 0.229 & 0.256 & 0.540 & 73.3 & 71.6 & 70.5 & $71.8 \pm 1.3$ & 1.79 & 0.231 & 0.261 & 0.549 & 73.8 & 73.0 & 71.7 & $72.9 \pm 1.1$\\
2.1 & 1.60 & 0.247 & 0.276 & 0.586 & 79.0 & 77.3 & 76.6 & $77.6 \pm 1.2$ & 1.79 & 0.250 & 0.287 & 0.606 & 80.0 & 80.1 & 79.1 & $79.8 \pm 0.5$\\
2.2 & 1.58 & 0.264 & 0.296 & 0.632 & 84.5 & 82.8 & 82.6 & $83.3 \pm 1.0$ & 1.77 & 0.258 & 0.299 & 0.648 & 82.5 & 83.7 & 84.6 & $83.7 \pm 1.1$\\
2.3 & 1.57 & 0.281 & 0.315 & 0.676 & 89.8 & 88.0 & 88.3 & $88.7 \pm 0.9$ & 1.79 & 0.267 & 0.311 & 0.678 & 85.3 & 86.9 & 88.5 & $86.9 \pm 1.6$\\
2.4 & 1.55 & 0.297 & 0.332 & 0.720 & 94.8 & 92.8 & 94.0 & $93.9 \pm 1.0$ & 1.79 & 0.299 & 0.344 & 0.738 & 95.6 & 96.3 & 96.4 & $96.2 \pm 0.4$\\
2.5 & 1.49 & 0.312 & 0.348 & 0.763 & 99.8 & 97.4 & 99.6 & $98.9 \pm 1.3$ & 1.78 & 0.311 & 0.357 & 0.759 & 99.4 & 99.7 & 99.1 & $99.5 \pm 0.3$\\
\hline
\end{tabular}
\end{center}
Notes --- notation and uncertainties are the same as those described in the
notes for Table~\ref{tab:EvidSPLE1D}.
\end{table*}


\subsubsection{The ACS posterior PDF for $\slope$}
\label{sec:LensModel:Result:slope}

In the previous section, we explored the \HST\ data constraints on the
slope parameter, optimizing the other parameters of the SPLE lens
model at each step.  To characterize properly $P(\slope|\dataVec,
\imModelSet=\imModelFive)$ in Equation
(\ref{eq:MargSlopePosterior:sec}), we would need to marginalize 
over all lens parameters $\lenspars$ instead.  However, as we shall
now see, this optimization approximation is actually a good one and is
certainly the most tractable solution due to the high dimensionality
of the problem (16 parameters to describe G1, G2 and external shear).
Direct sampling in the 16-dimensional parameter space
of $P(\dataVec | \slope, \lenspars,
\imModelSet=\imModelFive) \, P_{\rm no\, ACS}(\slope)\, P(\lenspars)$ in Equation
(\ref{eq:MargSlopePosterior:sec}) via, for example, Markov chain Monte
Carlo (MCMC) techniques using the extended source information is not
feasible on a reasonable time scale.  Importance sampling of the prior
PDF from the radio data of image positions and fluxes ($P_{\rm no\,ACS}(\slope,
\lenspars) = P_{\rm no\,ACS}(\slope, \lenspars | {\rm radio})$) by weighing the
samples by $P(\dataVec | \slope, \lenspars, \imModelSet=\imModelFive)$
is difficult since $\slope$ is effectively unconstrained by the radio
data (the $\chi^2$ changes by $\lesssim 1$ in the slope range between 
1.5 and 2.5).\footnote{We set $\slope_{\rm G2}=\slope_{\rm G1}=\slope$ since
  the slope of G2 is ill-constrained \citep{KoopmansEtal03}.}

It is precisely the unconstrained nature of the $\slope$ parameter that makes
the optimization approximation so good. 
The ``tube'' of
$\slope$-degeneracy traversing the 16-dimensional parameter space dominates
the uncertainties in the parameters.  We thus
assume that the tube of
$\slope$-degeneracy has negligible thickness (a degeneracy curve), 
and use $P(\dataVec | \slope, \lenspars,
\imModelSet=\imModelFive)$ to break the degeneracy.  Specifically, we
use the radio observations, \HST\ Near Infrared Camera and
  Multi-Object Spectrometer 1 (NICMOS) images (Proposal 7422;
  PI:Readhead), and time delay data to
obtain the best-fitting $\hat{\lenspars}$ for a given
$\slope$=$\slope_i$ (assuming $\OM=0.3$, $\OL=0.7$ and $w=-1$ in using
the time delay data), and compute the corresponding $P(\dataVec |
\slope_i, \hat{\lenspars}, \imModelSet=\imModelFive)$.  These are the
listed evidence values in the second column of Table
\ref{tab:slopeEvid} for the various $\slope_i$ values.  The time delay
data are included because the predicted relative Fermat potential
among the image pairs using the radio and NICMOS data are otherwise
inconsistent with one another.  The optimized parameters from only the 
radio and NICMOS data lead to $\chi^2\sim600$ for just the time delay 
data; including the time delay data reduces the time delay $\chi^2$ to
$\sim 1$ with only a mild increase in the radio and NICMOS $\chi^2$ of
$\sim 6$.  We
``undo'' the inclusion of the time delay data (so that we do not use
the time delay data twice in the importance sampling of Equation
(\ref{eq:CosmoparsPosteriorFullSimp2:sec})) by subtracting the log
likelihood of the time delay from the log likelihood of $\dataVec$;
the effect is negligible since the latter is $\sim10^4$ higher in
magnitude.
 
Our thin degeneracy tube assumption implies that 
$P(\dataVec | \slope) \simeq P(\dataVec | \slope, \hat{\lenspars})$, such that  
the posterior PDF for the slope is
$P(\slope | \dataVec) \propto P(\dataVec | \slope)\ P_{\rm no\,ACS}(\slope)$.
Assigning a uniform prior (i.e., $P_{\rm no\,ACS}(\slope)$ is constant),
we arrive at the result that our desired PDF is
just the exponentiation of the log evidence in column 2 of 
Table~\ref{tab:slopeEvid}.
Fitting these log evidences with the following quadratic function,
\be
\label{eq:logPfit}
\log P(\slope|\dataVec) = C - \frac{(\slope - \slope_0)^2}{2\sigma_{\slope}^2},
\ee
we obtain the following best-fit parameter
values: $\slope_0=2.081\pm0.027$,
$\sigma_{\slope}=0.0091\pm0.0008$, and $C=(1.60\pm0.01)\times10^4$.
While the PDF width 
$\sigma_{\slope}$ is very small, the centroid is not well
determined.  Adding $\sigma_{\slope}$ and the uncertainty in
$\slope_0$ in quadrature, we finally approximate $P(\slope | \dataVec)$
with a Gaussian centered on $2.08$ with
standard deviation $0.03$.  This provides the prior on $\slope$ from
the ACS data (in Equation (\ref{eq:CosmoparsPosteriorFullSimp2:sec})).

The deep ACS data therefore allow a significant improvement to the
previous measurement in \citet{KoopmansEtal03} of $\slope =
1.99\pm0.20$, which was based on the radio data and the NICMOS ring.
Coincidentally, our $\slope=2.08\pm0.03$ is identical, apart from the
spread, to the measurement from SLACS of
$\slope=2.08\pm0.2$ that was based on a sample of massive elliptical
lenses \citep{KoopmansEtal09}.  
The spread of 0.2 in the SLACS measurement is the intrinsic scatter of
slope values in the sample, and is larger than the typical
uncertainties associated with individual systems in the sample of
$\sim 0.15$.
We note that our measurement 
is not the first percent-level determination of a
strong lens density profile slope. \citet{WucknitzEtal04} used high
precision astrometric measurements from VLBI data to constrain the
$\slope$ parameter in B0218$+$357 to be $1.96 \pm 0.02$ (where we have
transformed their $\beta$ into our notation). However, they did not use
exactly the same model as we do here (instead working with
combinations of isothermal elliptical potentials and neglecting
external convergence).  \citet{DyeWarren05} measured the power-law
slope of the lens galaxy in the Einstein ring system 0047-2808 to be
$\slope = 2.11\pm0.04$ based on the extended image constraints.  More
recently, \citet{DyeEtal08} determined the power-law slope of the
extremely massive and luminous lens galaxy in the Cosmic Horseshoe
Einstein ring system J1004+4112 to be $\slope=1.96\pm0.02$.  


\subsubsection{Predicted relative Fermat potentials}

In order to be able to calculate the time delay likelihood function,
$P(\tdelayVec | \zd,\zs, \cosmopars, \slope,\kext,\imModelSet)$, at any value
of the slope~$\slope$, we need to interpolate the Fermat potential differences
given in Table~\ref{tab:slopeEvid}. In fact, these data give us the 
function $q(\slope)$ to insert into
Equation~(\ref{eq:fprelation:sec}): we can do the interpolation at $\kext = 0.0$
and then rescale by $(1-\kext)$ without loss of generality.

For each of the image pairs, we fit the relative Fermat potential difference
as a third-order polynomial function of $\slope$ using the values we have at
the discrete points $\slope_i$ for the SPLE models in the table.  Recall that
the SPLE model provides an unbiased estimate of the relative Fermat potential,
and that the various top data models $\imModelSet$ gave consistent estimates. 
Thus, the polynomial fit gives the function $q(\slope,\imModelSet)$ in
Equation (\ref{eq:fprelation:sec}).  The third-order polynomial fit leads to
residuals ($=(\fpdiff_i-\fpdiff^{\rm poly})/(\fpdiff_i)$) of $<1\%$ for all
image pairs at all slope points in Table \ref{tab:slopeEvid} except for
$\slope_i=1.7$, which has residuals of $\sim2\%$.


\section{Breaking the Mass-Sheet Degeneracy: Stellar Dynamics}
\label{sec:StellDyn}

In this section, we present the observations and data reduction for measuring
the velocity dispersion~$\sigma$ of G1 in \ourlens. This measurement appears
as the likelihood function given in Equation~(\ref{eq:vdispLikelihood:sec})
above.


\subsection{Observations}
\label{sec:StellDyn:Obs}

We have obtained a high signal-to-noise spectrum of \ourlens\ using
the Low-Resolution Imaging Spectrometer 
(LRIS; \citeauthor{OkeEtal95}~\citeyear{OkeEtal95})
on Keck 1. The data were obtained from the red side of the
spectrograph on 12 June 2007 using the 831/8200 grating with the D680 dichroic
in place. A slit mask was employed to obtain simultaneously spectra for two
additional strong lenses in the field \citep{FassnachtEtal06p2} and to
continue to probe the structure along the line of sight to the lens
\citep{FassnachtEtal06}. The night was clear with a nominal seeing of
0\farcs9, and 10 exposures of 1800s and one exposure of 600s were obtained for
a total exposure time of 18600s.

Each exposure was reduced individually using a custom pipeline \citep[see][for
details]{AugerEtal08} that performs a single resampling of the spectra onto a
constant wavelength grid; the same wavelength grid was used for all exposures
to avoid resampling the spectra when combining them, and an output pixel scale
of 0.915 \AA\ was used to match the dispersion of the 831/8200 grating.
Individual spectra were extracted from an aperture 0\farcs84 wide
(corresponding to 4 pixels on the LRIS red side) centered on the peak of the
flux of the lensing galaxy G1. The size of the aperture was chosen to avoid
contamination from the spectrum of G2 while maximizing the total flux for an
improved signal-to-noise ratio. The extracted spectra were combined by
clipping the extreme points at each wavelength and taking the
variance-weighted sum of the remaining data points. The same extraction and
coaddition scheme was performed for a sky aperture to determine the resolution
of the output co-added spectrum; we find the resolution to be ${\rm R} =
2560$, corresponding to $\sigma_{\rm obs} = 49.7\, {\rm km\, s^{-1}}$. The
signal-to-noise ratio per pixel of the final spectrum is $\sim 60$.


\subsection{Velocity dispersion measurement}
\label{sec:StellDyn:Model}

We use a Python-based implementation of the velocity-dispersion code from
\citet{vanderMarel94}, with one important modification. Our implementation
allows for a linear sum of template spectra to be modeled using a bounded
variable least squares solver with the constraint that each template must have
a non-negative coefficient. We use a set of templates from the INDO-US stellar
library containing spectra for a set of seven K and G giants with a variety of
temperatures and spectra for an F2 and an A0 giant. These templates of
early-type stars are particularly important for \ourlens, which has a
post-starburst spectrum \citep{MyersEtal95}.

\begin{figure}
\centering\includegraphics[width=75mm,clip]{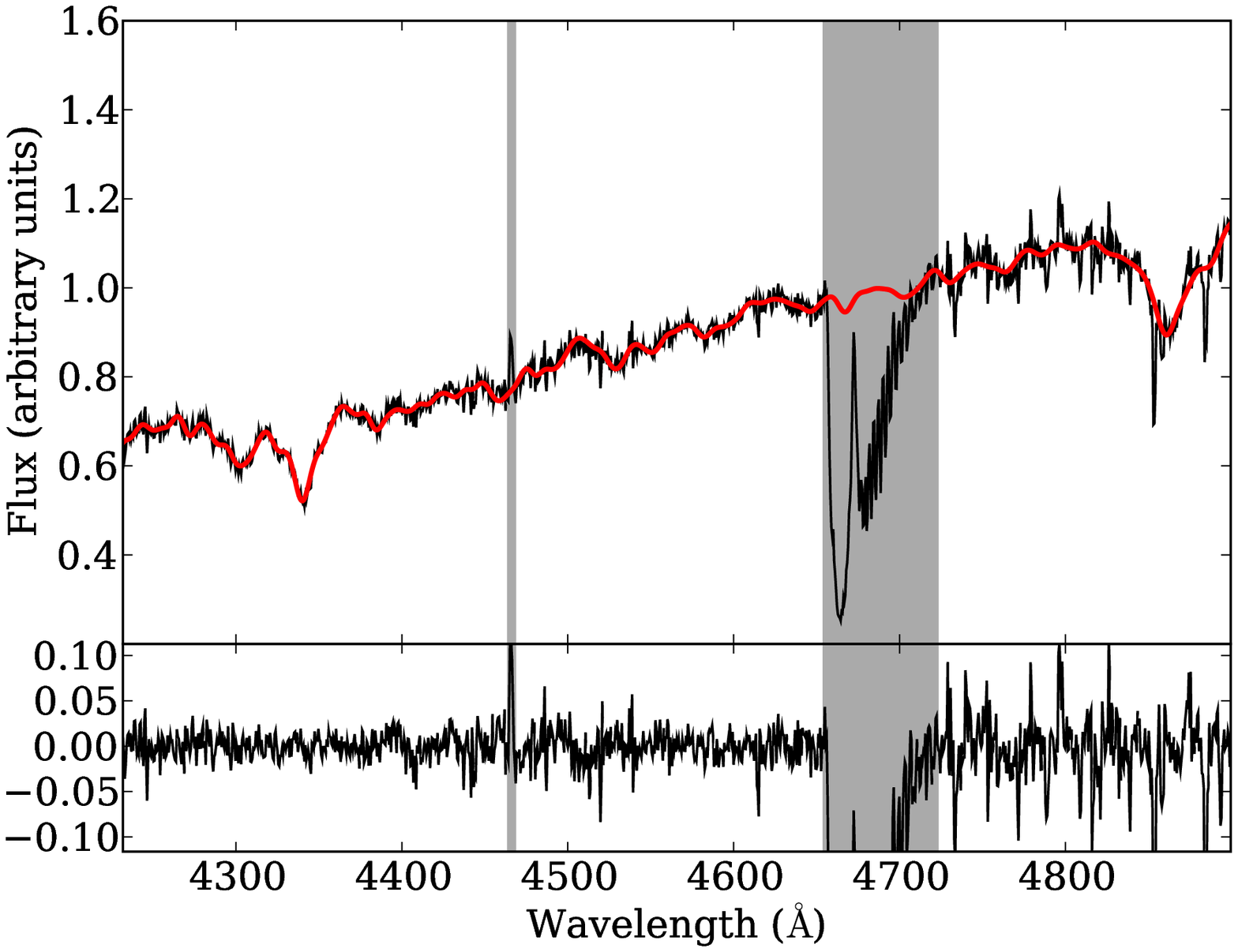}
\caption{\label{fig:LRISvelocitydispersion}  The LRIS spectrum of \ourlens\
(black line) with a model generated from all 9 INDO-US templates and
a 9th order continuum overplotted (red line).
The gray shaded areas were not included in the fit, and the lower panel shows
the fit residuals. The spectrum and our modeling suggest a central velocity
dispersion of $\sigma = 260 \pm 15\, {\rm km\, s^{-1}}$, including systematic
errors. }
\end{figure}

We perform our modeling over a wide range of wavelength intervals and find a
stable solution over a variety of spectral features; we therefore choose to use
the rest-frame range from 4200~\AA\ to 4900~\AA\ for our fit. The INDO-US
templates have a constant-wavelength resolution of 1.2~\AA\ which
corresponds to $\sigma_{\rm template} = 33.6\kms$ over this wavelength range.
We iterate over a range of template combinations and polynomial continuum
orders and find that a variety of solutions that vary around 
$260\, {\rm km\, s^{-1}}$ with a spread of about $13\, {\rm km\, s^{-1}}$ and
statistical uncertainties of $7.7\, {\rm km\, s^{-1}}$ (see Figure
\ref{fig:LRISvelocitydispersion}). We therefore adopt a
velocity dispersion of $\sigma = 260 \pm 15 \kms$, with the error
incorporating the systematic template mismatch and the statistical error for
the models.  This agrees with the previous measurement of $\sigma_{\rm
  ap}=247\pm35 \kms$ by \citet{KoopmansEtal03} with a significant
reduction in the uncertainties, though we note that the two velocity
dispersions have been measured in slightly different apertures.

\section{Breaking the Mass-Sheet Degeneracy: Lens Environment}
\label{sec:LensEnv}

In this section, we outline two approaches for quantifying
the prior probability distributions of the external mass sheet~$\kext$.
Computing this quantity such that  Equation (\ref{eq:MassSheet:H0bias}) holds true
is not a trivial matter. 
The non-linearity of strong lensing means that the
surface mass density at a given angular position in successive redshift planes
between the  observer and the source cannot simply be scaled by the appropriate 
distance ratios and summed: rather, the deflection angles (which can be large)
need to be taken into account when calculating the distortion matrices (which
contain and define the external convergence and shear),
leading us towards a ray-tracing approach
\citep{HilbertEtal09}. Detailed investigation of the ray paths down the
\ourlens\ light cone is beyond the scope of this paper, and we defer it to a
later work (Blandford et al.~in preparation). In this section we
use the statistics of \ourlens-like fields in numerical simulations to derive a
PDF for~$\kext$.


\subsection{Ray-tracing through the Millennium Simulation}
\label{sec:LensEnv:MS}

Following
\citet{HilbertEtal07}, we use the multiple-lens-plane algorithm to
trace rays through the Millennium Simulation
\citep[MS;][]{SpringelEtal05}, one of the largest N-body simulations
of cosmic structure formation.\footnote{ The details of the
  ray-tracing algorithm are described in \citet{HilbertEtal09}. The
  methods for sampling lines of sight, identifying strong lensing
  events, and calculating the convergence are described in
  \citet{HilbertEtal07}. Note that we also include a stellar component
  in the ray-tracing as described in \citet{HilbertEtal08}.
} 
We then identify lines of sight where strong lensing by matter
structures at $\zd=0.63$ occurs for sources at $\zs=1.39$.  The
convergence along these lines of sight is estimated by summing the
projected matter density on the lens planes weighted for a source at $\zs=1.39$$^9$ 
along the ray trajectory.
By excluding the primary lens plane at
$\zd=0.63$ that causes the strong lensing, the constructed convergence
is truly external to the lens and is due to the line-of-sight
contributions only. By sampling many lines of sight, we obtain an
estimate for the probability density function of $\kext$ from
simulations.  We denote this as the ``MS'' prior on $\kext$.

\begin{figure}
\includegraphics[width=75mm]{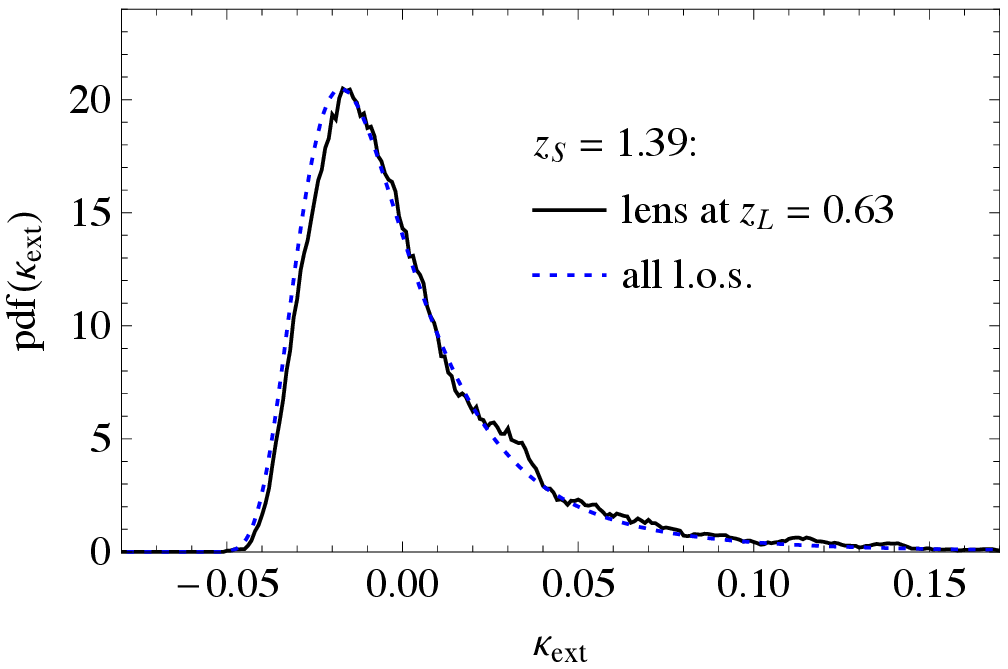}
\caption{
\label{fig:HilbertKext}
Probability distribution for the external convergence $\kappa_\mathrm{ext}$
along strongly lensed lines of sight from the Millennium Simulation for the
lens redshift $z_\mathrm{L}$ and source redshifts $z_\mathrm{S}$ of \ourlens{}
(solid line) compared to the convergence distribution for all lines of sight
(dotted line).
}
\end{figure}

Figure \ref{fig:HilbertKext} shows the predicted 
amount of external convergence constructed using 
$6.4 \times 10^8$ lines of sight (with and without strong
lenses) to sources at $\zs=1.39$: of these, $8.0\times 10^3$ lines
of sight contain strong lenses.
For both curves, the mean $\kext$ 
is consistent with zero with a spread of $\sim 0.04$.

How should we interpret this distribution? According to its 
definition, $\kext$ {\it could} have
contributions from galaxies on the primary lens plane that do not affect the
dynamics. Neglecting these contributions (effectively assuming that the lens
is an isolated galaxy) might lead to an underestimate of $\kext$, since most
lenses are massive galaxies that often live in over-dense environments like
galaxy groups and
clusters.\footnote{
\label{ftnote:MS_no_splitting}
It is beyond the scope of this paper to quantify this contribution from our
ray-tracing simulations. This would require modeling the lenses and their
environment in a way that allows one to split the mass distribution into a
part that is accounted for by the lens model (and constrained by lensing and
dynamics data) and a part that acts as external convergence.
} 
However, if the local contribution to the external convergence is accounted
for in the lensing plus dynamics modeling \citep[as discussed
in][]{FassnachtEtal06}, then the MS PDF will give an accurate uncertainty in
the inferred Hubble constant after marginalization.

Indeed, what the MS PDF also verifies is that {\it on average} the contribution
to the external convergence at a strong lens from line-of-sight structures is
almost the same as that for a random line of sight, namely zero.  
The
MS prior therefore suggests that ensembles of {\it isolated} strong lenses
will yield estimates of cosmological parameters that are not strongly biased 
by line-of-sight structures.
The PDF in Figure~\ref{fig:HilbertKext} gives us an idea of by how much
individual lenses' line-of-sight $\kext$ values vary, and hence an  estimate
of the uncertainty on $H_0$ due to this structure. In the absence of any other
information, we can assign  the Millennium Simulation PDF as a prior
on~$\kext$ in order to limit the possible values of external convergence to
those likely to occur. This assignment 
has the effect of adding an additional 
uncertainty of~$\sim 0.04$ in~$\kext$, with no systematic shift in~$\kext$. 


\subsection{Combining galaxy density observations with ray-tracing
simulations} 
\label{sec:LensEnv:OBS}

The prior discussed in the preceding section does not take into account any
information about the environment of \ourlens. Here, we combine knowledge of
the lens environment with ray-tracing to obtain a more informative prior on
the external convergence. 

\citet[][]{FassnachtEtal09} compared galaxy number counts in fields
around strong galaxy lenses, including \ourlens, with number counts in
random fields and in the COSMOS field. Among other measures, they used
the number of galaxies with apparent magnitude $18.5 \le
m_\mathrm{F814W} < 24.5$ in the F814W filter band in apertures of
$45\,\arcsec$ radius 
(300~kpc at the redshift of \ourlens) 
to quantify the galaxy number density
$n_\mathrm{gal}$ projected along lines of sight.  They found that the
distribution of $n_\mathrm{gal}$ for lines of sight containing strong
lenses is not very different from that for random lines of
sight. However, \ourlens\ lies along a line of sight with a galaxy
density $n_\mathrm{gal}$ that is about twice the mean over random
lines of sight, $\langle n_\mathrm{gal} \rangle$.
A positive $\kext$ bias can arise through Poissonian fluctuations that are
present in the number of  groups along the line of sight in the {\it observed}
sample of strong lenses. 

\begin{figure}
\includegraphics[width=75mm]{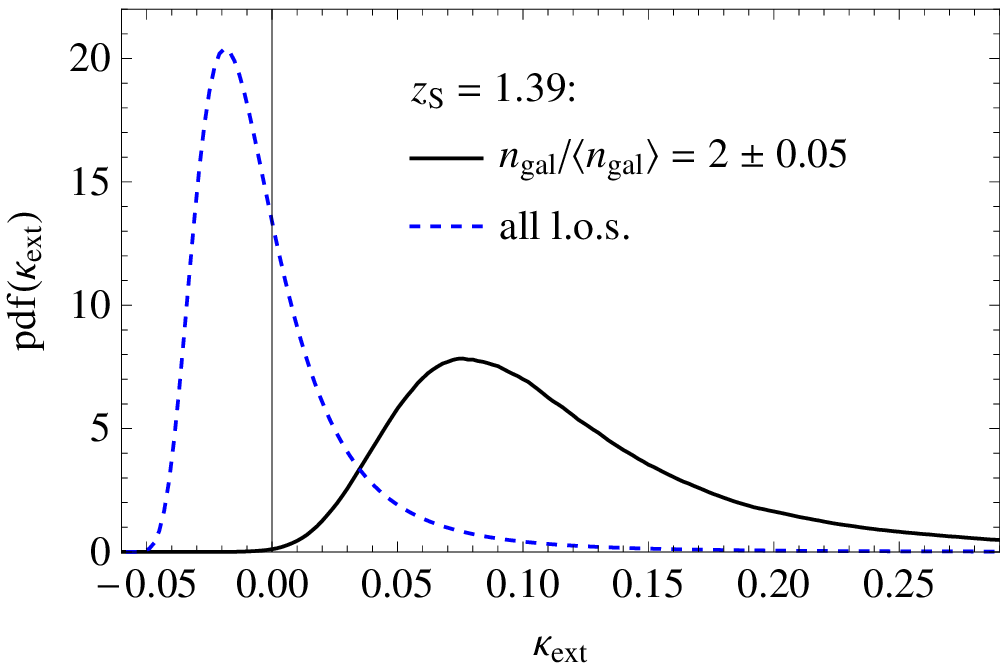}
\caption{\label{fig:HilbertKext2}
Probability distribution for the external convergence  $\kappa_\mathrm{ext}$
obtained from combining results of galaxy number counts around \ourlens\ with
results from ray-tracing through  the Millennium Simulation. Compared are the
distribution along lines of sight with a relative galaxy number density
$n_\mathrm{gal}/\langle n_\mathrm{gal}\rangle = 2.00\pm 0.05$ (solid line) to
the distribution along all lines of sight (dotted line).
}
\end{figure}

We can use this measurement of galaxy number density in the \ourlens\ field to
generate a more informative prior PDF for~$\kext$.
As for the MS prior in the previous section,
we use the ray-tracing through the MS together with the semi-analytic
galaxy model of \citet{DeLuciaBlaizot2007} to quantify the expected
external convergence $\kext$ for lines of sight with a
given {\it relative overdensity} $n_\mathrm{gal}/\langle n_\mathrm{gal}
\rangle$. 
Dividing out the absolute number of galaxies in the field
accounts for differences due to the particular set of
cosmological parameters used by the Millennium
Simulation and inaccuracies in the galaxy model: We assume that differences in
the relative 
overdensity between the MS cosmology and the true one are small.

We generate 32 simulated fields of
$4\times4\,\mathrm{deg}^2$ on the sky containing the positions and
apparent magnitudes\footnote{
The model galaxy catalogs do not provide
F814W magnitudes. We simply approximate $m_\mathrm{F814W}$ by
combining SDSS $i$-band and $z$-band magnitudes to get
$m_\mathrm{F814W} = x_i m_i + (1-x_i) m_z$ with $x_i = 0.5$.
We have checked that our results do not depend strongly on $x_i\in[0,1]$.
} of the model
galaxies at redshifts $0<z<5.2$ together with maps of the convergence
$\kappa$ to source redshift $\zs=1.39$. The galaxy positions
and magnitudes in the simulated fields are converted into maps of the
galaxy density $n_\mathrm{gal}$. We then select all lines of sight
with relative overdensity $1.95 \le n_\mathrm{gal}/\langle
n_\mathrm{gal} \rangle < 2.05$ and compute the distribution of the
convergence along these lines of sight. The resulting convergence
distribution (shown in Figure~\ref{fig:HilbertKext2}) is then used as
prior distribution for the external convergence $\kappa_\mathrm{ext}$,
which we denote as the ``OBS'' (observations and MS) prior.

The convergence computed in this way is not strictly speaking 
external convergence, since (i) we do not subtract any contribution from any primary strong lens, 
(ii) we take all lines of sight and not just those to
strong lenses. We are instead building on one of the results 
of the previous section and assume that the distribution of external convergences is
very similar to the distribution of convergences along random lines of sight.

Where this approach becomes inappropriate is where a ray passes close to a galaxy
center, and is hence associated with a very large convergence. Assuming such a line of sight
as foreground/background for a strong lens galaxy essentially creates a lens system
 with two or more strong deflectors.
These sightlines correspond to compound lenses such as {SDSS\ J0946+1006}
\citep{Gav++08}, but not to \ourlens.
However, the tail of high convergence values does not pose a problem here: 
as we will see in Section~\ref{sec:H0:nuisance}
below, the high external convergence is rejected by the
dynamics modeling. We expect the mean and width
of the PDF in Figure~\ref{fig:HilbertKext2} to represent well the possible
values of $\kext$ for a field that is over-dense in galaxy number by a factor
of two.

Our OBS $\kext$ distribution agrees with earlier estimates from
\citet{FassnachtEtal06}, who identified and modeled the 4 groups along
the line of sight to \ourlens\ using various mass assignment recipes.
In both approaches, we and \citet{FassnachtEtal06} are concerned
primarily with extracting information on the external convergence and
not the external shear.  
If we were to estimate the external convergence by assigning masses 
and redshifts to
all objects in the \ourlens\ field, and then ray tracing through the resulting
model mass distribution, the external shear as required in the strong lens
modeling would serve as an important calibrator for the external convergence.
Such a procedure is beyond the scope of this
paper, and we defer it to a future publication (Blandford et al., in
preparation).  However, we do find 
(by computing the distribution of external shears
in MS fields with different external convergences) that 
the magnitude of the external shear required by the
strong lens modeling ($\gamma_{\rm ext} \simeq 0.075)$ is consistent
with the external shear amplitude predicted in the OBS scenario for the 
\ourlens\ field.


\subsection{The influence on lens modeling.} 
\label{sec:Lensmodel:OBS}

As already remarked, the description of ray propagation in an inhomogeneous
cosmology is quite subtle. The matter (dark plus baryonic) density is
partitioned between virialized structures (galaxies, groups and clusters) and
a depleted background medium. Any structures sufficiently close to the line of
sight will imprint convergence and shear onto a ray congruence. Meanwhile the
background medium will contribute less Ricci focusing than would be present in
a homogeneous, flat universe and will diminish the net convergence. 

As the foregoing discussion makes clear, the line of sight to \ourlens\ is
unusual and we know quite a lot about the photometry and redshifts of the
intervening galaxies. It is therefore possible, in principle, to make a
refined estimate of the external convergence  and shear and to compare the
former with the simulations discussed above and the latter with the shear
inferred in the lens model described in \paperI. In this way, the shear, again
in principle, can be used to calibrate $\kext$.

There is a second complication that must be addressed. Matter inhomogeneities
in front of G1 and G2 distort the image of
the primary lens as well as the multiple images of the source. Inhomogeneities
behind the lens contribute further distortion in the images of the source.
In a more accurate approach, these effects should be taken into account explicitly in the construction of the lens model, while 
here we are subsuming them in a single correction factor $\kext$. The way that
the resulting corrections affect the inference of a value for $H_0$ turns out to be
quite complex. However, it appears that in
the particular case of \ourlens, the error that is incurred does not
contribute significantly to our quoted errors. 

These matters will be discussed in a forthcoming publication.


\section{Priors for model parameters $\pars$}
\label{sec:Priors}

A key goal of this work is to quantify the impact of the most serious
systematic errors associated with using time-delay lenses for cosmography. Our
approach is to characterize these errors as nuisance parameters, and then
investigate the effects of various choices of prior PDF on the inference of
cosmological parameters. To this end,  we use either well motivated priors
based on the results of Section \ref{sec:LensModel}, Section 
\ref{sec:LensEnv} and other independent studies, or, for contrast, 
uniform (maximally ignorant) prior PDFs. 
We now describe our choices for each parameter in turn.

\begin{itemize}


\item $P(\cosmopars)$.  We consider a set of four cosmological parameters,
  $\cosmopars = \{H_0, \OM, \OL, w\}$. We then assign the following 
  four different joint prior PDFs:
  \begin{description}
  \item{\bf K03:} uniform prior on $H_0$ between $0$ and $150\ \kmsMpc$,
    $\OM=0.3$, $\OL=0.7$, and $w=-1$.  This is the cosmology that was
    assumed in \citet{KoopmansEtal03} (the most recent $H_0$
    measurement from \ourlens\ before this work), and is the cosmology
    that is typically assumed in the literature for measuring $H_0$
    from time-delay lenses.  This form of prior allows us to compare
    our $H_0$ to earlier work.
  \item{\bf UNIFORM} priors on all four cosmological parameters, with
    either the $w=-1$ or the flatness ($\OM = 1 - \OL$) constraint
    imposed.  These priors allow us to
    quantify the information in the \ourlens\ data set as
    conservatively as possible. 
  \item{\bf WMAP5:} WMAP 5 year data set posterior PDF for $\{H_0,
    \OM, \OL, w\}$, assuming either $w=-1$ or a flat geometry.  This
    allows us to constrain either flatness or $w$ 
    by combining \ourlens\ with WMAP.
  \item{\bf WBS:} Joint posterior PDF for 
    $\{H_0, \OL, w\}$ with a flat geometry, given the WMAP5 data in combination
    with compendia of BAO and supernovae (SN) data
    sets.  This allows us to quantify the
    gain in precision made when incorporating \ourlens\ into the current global
    analysis. 
  \end{description}
The last two priors are defined by the Markov chains provided by the WMAP
team\footnote{\texttt{http://lambda.gsfc.nasa.gov}} based on the analysis 
performed by \citet{DunkleyEtal09} and \citet{KomatsuEtal09}.
The BAO data incorporated were taken from \citet{PercivalEtal07}; the
SN sample used is the ``union'' sample of
\citet{KowalskiEtal08}. While the BAO and SN data sets are
continually improving \citep[e.g.][]{HickenEtAl09}, this particular
well-defined snapshot is sufficient for
us to explore the
relative information content of our data set compared with other, well-known
cosmological data sets. We also note that 
the publication of Markov chain representations of posterior PDFs 
makes further joint analyses like the one we present here very
straightforward indeed.


\item $P(\slope)$.  We
  consider three different prior PDFs for the density profile slope.  
  In the first two priors, we
  ignore the \ourlens\ ACS data (i.e., dropping $P(\dataVec | \slope,
  \lenspars, \imModelSet=\imModelFive)$ in 
  Equation (\ref{eq:MargSlopePosterior:sec})); these first two are controls,  
  to allow the assessment of the amount of
  information contained in the ACS data.
  \begin{description}
  \item{\bf Uniform:} a maximally ignorant prior PDF, defined in the range 
    $1.5 \leq \slope \leq 2.5$. 
  \item{\bf SLACS:} This is a Gaussian prior based on the 
    result from the SLACS project:
    $\slope=2.08\pm0.2$ 
    \citep{KoopmansEtal09}.
    This was derived from a sample of low-redshift massive elliptical lenses,
    studied with combined strong lens and stellar dynamics modeling.
    We note that this was obtained without considering the presence of
    any external convergence $\kext$.  However, \citet{TreuEtal09} find
    that the environmental effects in the SLACS lenses are smaller than
    their measurement errors and are typically undetected.  Since SLACS
    lenses do not require an external shear in the modeling, typical
    $\kext$ values for these lenses are expected to be small.  Only in a
    few extreme cases does the $\kext$ reach values of order
    $0.05$--$0.10$. Therefore, we take directly the prior on the slope
    from SLACS lenses without corrections for $\kext$.
  \item{\bf ACS:} This prior is the PDF
    $P(\slope | \dataVec, \imModelSet=\imModelFive)$ obtained from the 
    analysis of
    the ACS image of \ourlens\ in Section~\ref{sec:LensModel:Result}.
    This is the most informative of the three priors on $\slope$, as it
    is determined directly from the B1608+656 data, independent of
    external priors from samples of galaxies (e.g.~SLACS).
  \end{description}


\item $P(\lenspars)$.  As described in
  Section~\ref{sec:LensModel:Result}, 
  we use the radio observations and the NICMOS F160W images 
  of \ourlens\ to constrain the smooth lens model parameters
  $\lenspars$ for a given slope
  $\slope$. The posterior PDF from this analysis forms the prior PDF for the
  current work.


\item $P(\kext)$.  We consider three forms of prior for the external convergence:
\begin{description}
\item{\bf Uniform} between $-0.25$ and $+0.25$: again, such a 
  maximally ignorant prior, again to provide contrast.
\item{\bf MS:} from the strong lenses in the MS, discussed in 
  Section~\ref{sec:LensEnv:MS}.
\item{\bf OBS:} from the galaxy number counts in the field of
  \ourlens\ and the MS, discussed in Section~\ref{sec:LensEnv:OBS}. 
\end{description}


\item $P(\aniso)$.  For the lens galaxy stellar orbit radial anisotropy
  parameter~$\aniso$, we simply assign a uniform prior between $0.5 r_{\rm
  eff}$ and $5 r_{\rm eff}$, where $r_{\rm eff}$ is the effective
radius that is determined from the photometry to be $0.58''\pm0.06''$ \citep{KoopmansEtal03} for
  the velocity dispersion measurement. The uncertainty in $r_{\rm
    eff}$ has negligible impact on the model velocity dispersion.
  The inner cutoff of $\aniso$ is motivated by
  observations \citep[e.g.,][]{KronawitterEtal00} and radial instability
  arguments \citep[e.g.,][]{MerrittAguilar85,StiavelliSparke91},
  while the outer cutoff is
  for computational simplicity (the model velocity dispersion changes by a
  negligible amount between $\aniso = 5 r_{\rm eff}$ and
  $\aniso\rightarrow\infty$).  These boundaries are consistent with
  those in \citet{GebhardtEtal03}.

\end{itemize}

These priors are summarized in Table~\ref{tab:priors}.


\section{Inference of $H_0$ and dark energy parameters from \ourlens}
\label{sec:H0}

In this section we present the results of the analysis outlined in
Section~\ref{sec:H0ProbTheory}, putting together all the likelihood 
functions and prior PDFs described in Sections
\ref{sec:LensModel} to \ref{sec:Priors}.  We obtain
$P(\cosmopars|\tdelayVec,\dataVec,\vdisp)$ by importance sampling, using the
two likelihoods in Equation~(\ref{eq:CosmoparsPosteriorFullSimp2:sec}) as the
weights for the
various priors on $\slope$, $\kext$, $\aniso$, and $\cosmopars$ listed in
Table~\ref{tab:priors} (see Appendix~\ref{app:H0ProbTheory:ImpSamp}
for details).  By using the likelihood functions of our \ourlens\
data sets, we are incorporating the uncertainties
associated with these measurements.
We expect and indeed find that the data are relatively 
insensitive to $\aniso$ and do not constrain it.
Focusing first on the systematic errors now quantified as the
nuisance
parameters $\slope$ and~$\kext$, we gradually increase the complexity of
the cosmological model to probe the full space of parameters. 

For each possible combination of the priors on the parameters in
Table~\ref{tab:priors}, we generate 96000 samples of $\slope$, $\kext$,
$\aniso$, and $\cosmopars$ to characterize the prior probability
distribution.   We also have two types of stellar distribution functions,
Hernquist and Jaffe, for modeling the stellar velocity dispersion;  we find
that the two different types of stellar distribution function produce nearly
identical PDFs for the cosmological parameters.  Since the priors on the
parameters play a greater role than does the choice of stellar dynamics model,
we focus only on the Hernquist stellar distribution function for the remainder
of the section.

\begin{table*}
\begin{center}
\caption{\label{tab:priors} Priors on the parameters}
\begin{tabular}{|l||c|c|c|}
\hline
$P(\slope)$ & uniform ($1.5\leq\slope\leq2.5$) & SLACS ($\slope=2.08\pm0.2$) & ACS ($\slope=2.08\pm0.03$) \\ 
\hline
$P(\kext)$ & uniform ($-0.25\leq\kext\leq0.25$) & MS (Millennium
Simulations; Figure \ref{fig:HilbertKext}) & OBS (Observations and MS; Figure
\ref{fig:HilbertKext2})  \\
\hline
$P(\aniso)$ & \multicolumn{3}{|c|}{uniform ($0.5r_{\rm eff}\leq\aniso\leq5 r_{\rm eff}$)}\\
\hline
$P(\cosmopars)$ & K03 ($\OM=0.3$, $\OL=0.7$, $w=-1$, & UNIFORMopen ($w=-1$, & UNIFORMw ($\OL=1-\OM$ uniform $\in [0,1]$, \\
& uniform $H_0 \in [0,150]\,\kmsMpc$) & $\OM$ and $\OL$ uniform $\in [0,1]$,  & uniform $w \in [-2.5,0.5]$, \\
&  & uniform $H_0 \in [0,150]\,\kmsMpc$) & uniform $H_0 \in [0,150]\,\kmsMpc$) \\
\cline{2-4}
& WMAPopen & WMAPw (WMAP5 with & WBSw (WMAP5 + BAO + SN with \\
& (WMAP5 with $w=-1$) & flatness and time-independent $w$) & flatness and time-independent $w$) \\
\hline
\end{tabular}
\end{center}
Notes --- The K03 entry for $P(\cosmopars)$ is the same prior as
in \citet{KoopmansEtal03}.  This is also the most common cosmology prior
assumed in previous studies of time-delay lenses.
\end{table*}


\subsection{Exploring the degeneracies among $H_0$, $\slope$ and~$\kext$}
\label{sec:H0:nuisance}

To investigate the impact of our limited knowledge of the lens density profile
slope~$\slope$ and external convergence~$\kext$, we first fix the cosmological
parameters $\OM$, $\OL$ and $w$ according to the K03 prior. This allows us a
simplified view of the problem, and also a comparison with previous work that
used this rather restrictive prior.

\begin{figure*}
\begin{minipage}{0.48\linewidth}
\centering\includegraphics[width=\linewidth]{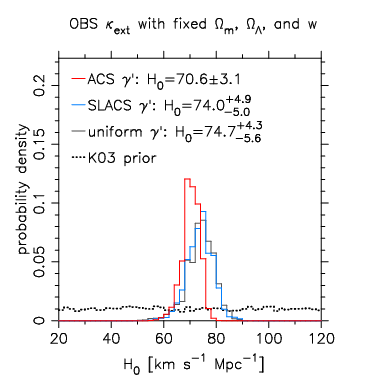}
\end{minipage}\hfill
\begin{minipage}{0.48\linewidth}
\centering\includegraphics[width=\linewidth]{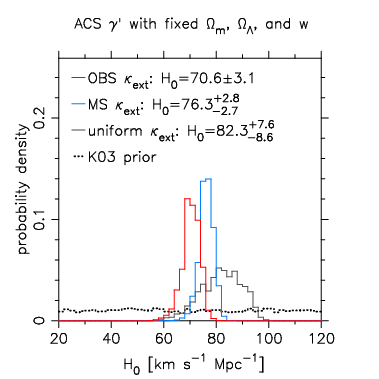}
\end{minipage}
\caption{\label{fig:H0-nuisance} Left: 
the marginalized posterior PDF for $H_0$
assuming K03 cosmology and OBS $\kext$ priors. 
Right: the marginalized posterior PDF for $H_0$
assuming K03 cosmology and ACS $\slope$ priors. The prior 
uncertainty in external convergence determines the precision of 
the inferred Hubble constant.}
\end{figure*}

We first assign the OBS prior for $\kext$, 
and look at the effect of the various
choices of density profile slope priors. The left-hand panel in 
Figure~\ref{fig:H0-nuisance} shows the
marginalized posterior PDF for $H_0$ for the three different priors for
$\slope$ given in Table~\ref{tab:priors}.  From this graph, we see
that the SLACS prior gives a similar estimate of $H_0$ as the uniform
prior with a negligible increase in precision.  
The ACS prior lowers $H_0$ relative to that of the SLACS
and uniform priors, and improves the precision in $H_0$ to $4.4\%$.  
Overall, the impact of the prior on $\slope$ is
relatively low in the sense that,
even with a uniform prior on~$\slope$, 
$H_0$ is still constrained to $7\%$ (taking
$H_0=\HK$ as our reference value).  
For the remainder of this paper, we assign the ACS prior.

As expected, the prior for $\kext$ has a greater effect, shown in
the right-hand panel of
Figure~\ref{fig:H0-nuisance}. Taking the maximally informative OBS prior as our
default, we see that relaxing this to the MS prior causes an increase in
inferred $H_0$ value of some $6\kmsMpc$, and relaxing further to a uniform
prior increases it by $12\kmsMpc$. The precision in $H_0$ also drops
by more than a factor of two from the OBS prior to the uniform prior.
Our knowledge of $\kext$ is therefore limiting the inference of $H_0$.

We note that the stellar dynamics contain a significant amount
of information on~$H_0$.  The stellar dynamics effectively constrain $\kext$
and $\slope$ to an approximately linear relation, where an increase in $\kext$
requires a steepening of the slope in order to keep the predicted velocity
dispersion the same.  Therefore, for a fixed range of $\slope$ values, 
the modeling of the stellar dynamics would only permit a corresponding
range of $\kext$ 
values.  Specifically, without dynamics as constraints, we find
$H_0=\HKnodyn^{\HKnodynhiOne}_{\HKnodynloOne}\kmsMpc$ for the ACS and
OBS priors. 
The lower bound on $H_0$ is somewhat weakened
by the high tail of the OBS $\kext$
distribution.  On the
other hand, this high tail is rejected by the use of
the dynamics data.  Therefore, our tight constraint on $H_0$ results 
from the \textit{combination} of 
all available data sets -- each data set constrains different parts
of the parameter space such that the joint distribution is tighter
than the individual ones.

To summarize, using all available information on \ourlens\ and the ACS
and OBS priors gives $H_0 = \HK\pm{\HKerrOne} \kmsMpc$, a precision 
of $4.4\%$.  We interpret
Figure~\ref{fig:H0-nuisance} as evidence that we are approaching saturation in the information we have on the lens model for \ourlens:
the mass model is now so well constrained that the inference of cosmological
parameters from this system is limited by our knowledge of the lens
environment. We now explore this joint inference in more detail, 
first putting it in some historical context.


\subsection{Comparison with other lensing $H_0$ results}
\label{sec:H0:litrev}

What improvement in the measurement of $H_0$ do we gain from our new
observations of \ourlens?  The most recent
measurement before this work by \citet{KoopmansEtal03} was $H_0=75^{+7}_{-6}
\kmsMpc$.  This result was based on a joint lensing and dynamics
modeling using the radio data, shape of the Einstein ring from the
NICMOS images and the earlier less precise velocity dispersion 
measurement.  Our improved analysis using the
deep ACS images and the newly measured velocity dispersion reduce the
uncertainty by more than a factor of two, even with the inclusion of the
systematic error due to the external convergence that was previously
neglected.  We attribute our lower $H_0$ value to our incorporation of
the realistically-skewed OBS $\kext$.

Let us now compare our $H_0$ measurement based on the K03 cosmology to several
recent measurements (within the past five years) from other time-delay
lenses.  Most analyses assumed $\OM=0.3$ and $\OL=0.7$ --- we point out
explicitly the few that did not.  
In B0218+357, \citet{WucknitzEtal04} measured $H_0=78 \pm
6\,\kmsMpc$ ($2 \sigma$) by modeling this two-image lens system with
isothermal elliptical potentials (and effectively measuring $\slope$, see
Section~\ref{sec:LensModel:Result:slope}) but neglecting external convergence.
\citet{YorkEtal05} refined this using the centroid position of the spiral
lens galaxy based on \HST/ACS observations as a constraint;
depending on the spiral arm masking, they found $H_0=70 \pm 5\,\kmsMpc$
(unmasked) and $H_0=61\pm7\,\kmsMpc$ (masked) (both with $2 \sigma$ errors).  
In the two-image {FBQ\,0951+2635},
\citet{JakobssonEtal05} obtained $H_0 = 60^{+9}_{-7}$ (random,
$1\sigma$) $\pm 2$ (systematic) $\kmsMpc$ for a singular isothermal
ellipsoid model and $H_0 = 63^{+9}_{-7}$ (random, $1 \sigma$) $\pm 1$
(systematic) $\kmsMpc$ for a constant mass-to-light ratio model,
again ignoring external convergence.
In the two-image quasar system SDSS\,J1650+4251, \citet{VuissozEtal07} found
$H_0=51.7^{+4.0}_{-3.0}\kmsMpc$ assuming a singular isothermal sphere and
constant external shear for the lens model.  More general lens models 
considered by these authors (e.g.\ 
including lens ellipticity, or using a de Vaucouleurs density profile) 
were found to be underconstrained.  
In the two-image quasar system SDSS\,J1206+4332,
\citet{ParaficzEtal09} found $H_0=73^{+3}_{-4}\kmsMpc$ using singular
isothermal ellipsoids or spheres to describe the three lens galaxies,
where photometry was used to place additional constraints on the lens
parameters.
Recently, \citet{FadelyEtal09} modeled the gravitational lens
Q0957+561 using four different dark matter density profiles, each with
a stellar component.  The lens is embedded in a cluster, and the
authors constrained the corresponding 
mass sheet using the results of a weak lensing analysis by
\citet{NakajimaEtal09}.  Assuming a flat universe with $\OM=0.274$ and
cosmological constant $\OL=0.726$, they found
$H_0=85^{+14}_{-13}\kmsMpc$, where the principle uncertainties were
due to the weakly constrained stellar mass-to-light ratio (a
manifestation of the radial profile degeneracy in the lens model).
Imposing constraints from stellar population synthesis models led to
$H_0=79.3^{+6.7}_{-8.5}\kmsMpc$.\footnote{\label{ftnote:Hdiffcosmo}
The corresponding $H_0$ for the K03 cosmology is within $\sim0.1\%$ of
the listed values.} 

In a nutshell, most of the recent
$H_0$ measurements from individual systems assumed isothermal
profiles, and neglected the effects of both $\slope$ and $\kext$:  
we interpret the significant variation
between the $H_0$ estimates in the recent literature as being due to these
model limitations. 
In contrast, our \ourlens\ analysis explicitly 
incorporates the uncertainties due to our lack of knowledge of
both $\slope$ and $\kext$. 
In fact, a spread of $\sim0.2$ in $\slope$ around
$2.0$ would give a spread of $\sim 40\%$ in $H_0$ for the cases where
isothermal lenses are assumed \citep{Wucknitz02}. 
These in turn are set by a lack of information on the
systems, either because only two images are formed, or the extended
source galaxy is not observed.

Other groups have looked to improve the constraints on $H_0$ by combining
several lenses together in a joint analysis.
Using a sample of 10 time-delay lenses, \citet{SahaEtal06} measured
$H_0=72^{+8}_{-11}\kmsMpc$ by modeling the lens' convergence distributions 
on a grid and using
the point image positions of the lenses as constraints 
(the PixeLens method).  \citet{Coles08} improved on the method and
obtained $H_0=71^{+6}_{-8}\kmsMpc$ while addressing more clearly their 
prior assumptions.
\citet{Oguri07} used a sample of 16 time-delay
lenses to constrain $H_0=68\pm6{\rm(stat.)}\pm 8 {\rm(syst.)}
\kmsMpc$ (for $\OM=0.24$ and $\OL=0.76$; see footnote
\ref{ftnote:Hdiffcosmo}) by employing a statistical approach based
on the image configurations.  
By simultaneously modeling
SDSS J1206+4332 with four other systems using PixeLens, \citet{ParaficzEtal09} 
derive $H_0 = 61.5^{+8}_{-4}\kmsMpc$.  
The larger quoted error bars on these ensemble estimates are perhaps a
reflection of the paucity of information available for each lens, as 
discussed above. All four analyses effectively assume that the ensemble 
external convergence
distribution has zero mean, which may not be accurate: for example, 
\citet{Oguri07} constructed a sample for which external 
convergence could be neglected, and then incorporated this into the systematic
error budget.  Furthermore, \citet{Oguri07} imposed a Gaussian prior on the
slope of $\slope=2.00\pm0.15$, and the PixeLens method's priors on
$\kappa$ may well implicitly impose constraints on $\gamma$ that are
similar to the prior in \citet{Oguri07}; these priors on the slope
may not be appropriate for individual systems in the ensembles.

In contrast, our measurement of $\slope$ from the ACS data means
that our results are independent of external priors on $\slope$.  In
fact, our detailed study of the single well-observed lens \ourlens,
even incorporating the effects of $\kext$, constrains $H_0$ better
than the studies using ensembles of lenses.  Our claim is that our
analysis of the systematic effects in \ourlens\ --- explicitly
including density profile slope and external convergence as nuisance
parameters --- is one of the most extensive on a single lens, and is
rewarded with one of the most accurate measurements of $H_0$ from
time-delay lenses.


\subsection{Relaxing the K03 prior}
\label{sec:H0:relaxcos}

\begin{figure}[!ht]
\centering\includegraphics[width=0.95\linewidth]{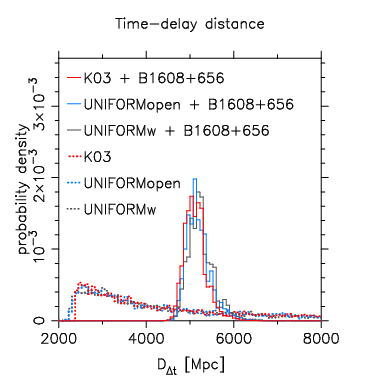}
\caption{\label{fig:Dtp} PDFs for $\tdist$, 
showing the \ourlens\ posterior 
constraints on $\tdist$ (solid) given 
assorted uniform priors for the cosmological parameters (dotted, labeled).
See the text for a full description of these various priors.
In this figure we assign the ACS and OBS priors for
$\slope$ and~$\kext$. \ourlens\ provides tight constraints on $\tdist$,
which translates into information about $\OM$, $\OL$ and
$w$ as well as $H_0$.}
\end{figure}

As we described in Section \ref{sec:H0theory},  strong lens time delays enable
a measurement of a cosmological distance-like quantity,  $\tdist \equiv
(1+\zd) D_{\rm d} D_{\rm s}/D_{\rm ds}$. While there is some slight further
dependence on cosmology in the  stellar dynamics modeling, we expect this
particular distance combination to be well constrained by the system. To
illustrate this,  we plot in  Figure~\ref{fig:Dtp} the PDF for $\tdist$ with
and without the constraints from \ourlens, for various choices of the cosmological
parameter prior PDF. Specifically, we show the effect of relaxing the prior on
$\OM$, $\OL$ and $w$ from the K03 delta function to the two types of uniform
distributions detailed in Table~\ref{tab:priors}: 
``UNIFORMopen'' and ``UNIFORMw''.
We see that all of these distributions
predict the same uninformative prior for $\tdist$, and that the \ourlens\
posterior PDFs are correspondingly similar. With the OBS and ACS priors for
$\kext$ and $\slope$, we estimate 
$\tdist \simeq \DtOne\, {\mathrm{Mpc}}$, a precision of
$\sim5\%$. 
The difference between the $\tdist$ estimates among the three priors
shown is $\lesssim2\%$.

Figure~\ref{fig:Dtp} suggests that a shifted log normal 
approximation (to take into account the skewness) 
for the product of the \ourlens\
likelihood function, marginalized over the OBS and ACS priors, 
is an appropriate compression of our results. We find that
\bea
\label{eq:DtpLikelihood}
\hspace{-0.3cm}\lefteqn{P(\tdist|H_0,\OM,\OL,w) \simeq }\nonumber\\
 && \hspace{-0.3cm} \frac{1}{\sqrt{2\pi} (x-\lambda_{\rm D}) \sigma_{\rm D}} 
\exp{\left[-\frac{(\log(x-\lambda_{\rm D}) - \mu_{\rm D})^2}{2\sigma_{\rm D}^2}\right]},
\eea
where $x=\tdist/(1\, {\rm Mpc})$, $\lambda_{\rm D} = \Dtfitlam$, $\mu_{\rm
  D}=\Dtfitmu$ and $\sigma_{\rm D} = \Dtfitsig$, accurately
reproduces the cosmological parameter inferences: for example,
Hubble's constant is recovered to $<0.7\%$ and its $16^{\rm th}$ and
$84^{\rm th}$ percentiles (68\% CL) are recovered to $<1.1\%$ for the
WMAP cosmologies we considered.


\subsection{Constraints on $\OM$ and $\OL$}
\label{sec:H0:opencos}

\begin{figure*}
\centering\includegraphics[height=0.95\linewidth,angle=270]{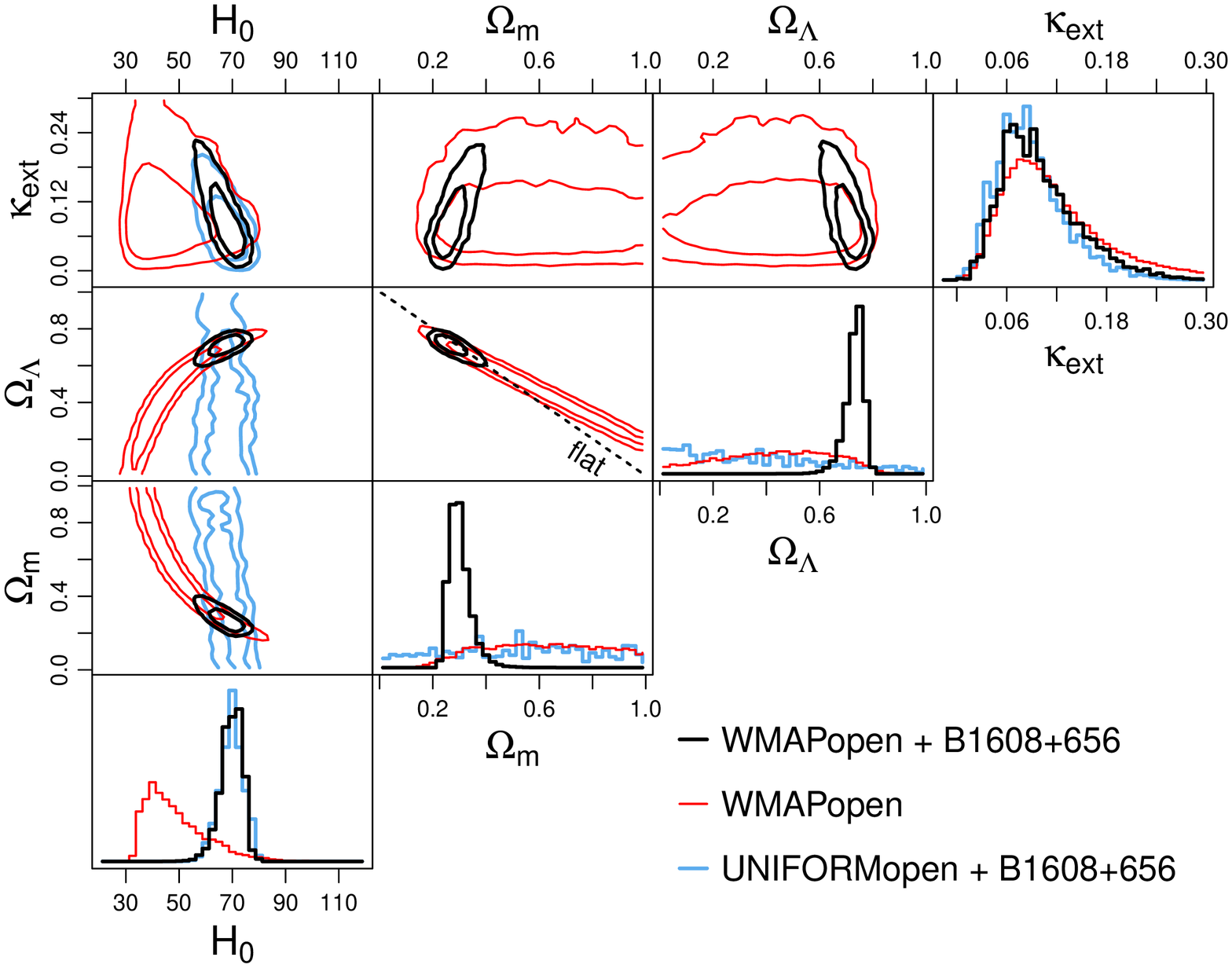}
\caption{\label{fig:cosuniform} The \ourlens\ marginalized posterior PDF for
  $H_0$, $\OM$, $\OL$ and $\kext$ in a $w=-1$ 
  cosmological model and assuming ACS $\slope$ and OBS
  $\kext$ priors; contours are 68\% and 95\% confidence levels. 
  The three sets of colored contours correspond to three different
  prior/data set combinations. 
  Blue: \ourlens\ constraints, given the UNIFORMopen prior; red: the prior
  provided by the WMAP 5 year data set alone; black: the joint constraints from
  combining WMAP and \ourlens.  The blue contours in the $\OM$ and
  $\OL$ columns are omitted since they would show almost no constraints,
  as indicated by the diagonal panels.}
\end{figure*}

Based on the construction of $\tdist$, we expect strong lens time delays
to be more  sensitive to $H_0$ than the other three cosmological parameters.
This is shown in Figure~\ref{fig:cosuniform}, where we consider 
$w=-1$ uniform cosmological prior ``UNIFORMopen'' 
and plot the marginalized \ourlens\ posterior PDF
to show the influence of the
lensing data (blue lines). While there is a slight dependence on $\OL$, we
see that the \ourlens\ data do indeed primarily constrain $H_0$. In contrast,
we plot the posterior PDF from the analysis of the 5 year WMAP data set
\citep[red lines,][]{DunkleyEtal09}. With no constraint on the curvature of
space, the CMB data provides only a weak prior on $H_0$, which is highly
degenerate with $\OM$ and 
$\OL$. Importance sampling the WMAP MCMC chains with the \ourlens\ likelihood,
we obtain the joint posterior PDF, plotted in black. 

Strong lens time delays are an example of a kinematic cosmological probe,
i.e., one that is sensitive to the geometry and expansion rate of the
Universe, but not to dynamical assumptions about the the growth of structure in the Universe. In
Table~\ref{tab:OkConstraint}, we compare the \ourlens\ data set to a number of
other kinematic probes from the literature. The WMAP data constrain the
angular diameter distance to the last scattering surface; these other data sets
effectively provide a second distance estimate that breaks the degeneracy
between $H_0$ and the curvature of space. In the \ourlens\ case, we constrain
$\Ok$  to be $\OkWO_{\OkWOloTwo}^{\OkWOhiTwo}$ (95\% CL).   We can see that in terms of
constraining the curvature parameter,  \ourlens\ is more informative than
the \HST\ Key Project $H_0$ measurement, and is comparable to the
current SNe Ia data set.

\begin{table}
\begin{center}
\caption{\label{tab:OkConstraint} Curvature parameter 
constraints from WMAP5 combined
with various data sets assuming $w=-1$ (95\% CL).}
\begin{tabular}{lcc}
\hline
WMAP5$^{\rm a,b}$             &           $-0.285 < \Ok < 0.010$ & $15\%$           \\
WMAP5 + \HST\, KP$^{\rm b,c}$ &           $-0.052 < \Ok < 0.013$ & $3.3\%$          \\ 
WMAP5 + SN$^{\rm b,d}$        &           $-0.032 < \Ok < 0.008$ & $2.0\%$          \\
WMAP5 + BAO$^{\rm b,e}$       &           $-0.017 < \Ok < 0.007$ & $1.2\%$          \\
{\bf WMAP5 + \ourlittlelens}  & $\mathbf{\OkWOTwo} $ & $\mathbf{2.0\%}$ \\
\hline
\end{tabular}
\end{center}
The third column 
gives the ``precision,'' 
quantified as half the 95\% confidence interval in $(1.0 - \Ok)$, 
as a percentage.$\;\;$
$^{\rm a}$ \texttt{http://lambda.gsfc.nasa.gov} $\;\;$
$^{\rm b}$ \citet{KomatsuEtal09}.$\;\;$
$^{\rm c}$ \citet{FreedmanEtal01}.$\;\;$
$^{\rm d}$ Based on the ``union'' SN samples compiled by \citet{KowalskiEtal08}.$\;\;$
$^{\rm e}$ \citet{PercivalEtal07}.$\;\;$
\end{table}

Figure~\ref{fig:cosuniform} also shows the primary nuisance parameter,
$\kext$. When \ourlens\ and the WMAP data are combined, the PDF for $\kext$
shifts and tightens very slightly, as we expect from the discussion in
Section~\ref{sec:H0:nuisance}.  If we relax the OBS prior on $\kext$ to 
uniform, then we obtain $-0.032<\Ok<0.021$ (95\% CL), which is still 
tighter than the \HST\ KP constraints.


\begin{figure*}[!ht]
\centering\includegraphics[height=0.95\linewidth,angle=270]{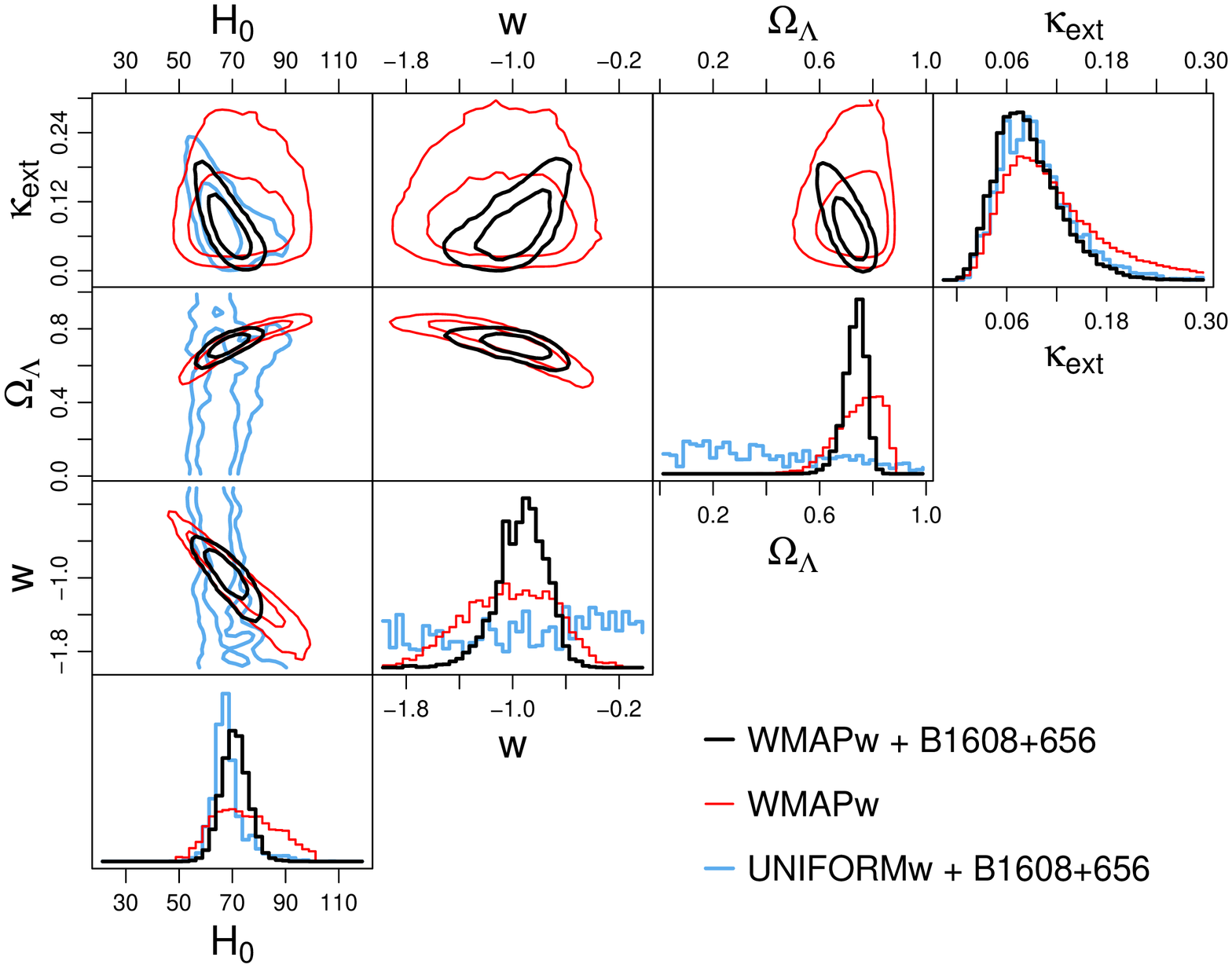}
\caption{\label{fig:wuniform} The \ourlens\ marginalized posterior PDF for
  $H_0$, $\OL$, $w$ and $\kext$ in a flat
  cosmological model, again assuming ACS $\slope$ and OBS
  $\kext$ priors; contours are 68\% and 95\% confidence levels. 
  The three sets of colored contours correspond to three different
  prior/data set combinations. 
  Blue: \ourlens\ constraints, given the UNIFORMw prior; red: the prior
  provided by the WMAP 5 year data set alone; black: the joint constraints from
  combining WMAP and \ourlens.  The blue contours in the $\OM$ and
  $\OL$ columns are omitted since they would show almost no constraints,
  as indicated by the diagonal panels.}
\end{figure*}

\subsection{Constraints on dark energy}
\label{sec:H0:DarkEnergy}

As noted by many authors \citep[e.g.][]{Hu05,KomatsuEtal09,RiessEtal09}, 
the degeneracy-breaking shown in the previous
subsection can be recast as a mechanism for constraining the equation of state
of dark energy,~$w$. 
If we assert a precisely flat geometry for the Universe,
as motivated by the inflationary scenario, we can spend our available
information on constraining~$w$ instead. Figure~\ref{fig:wuniform} shows the
marginalized posterior PDF for the cosmological parameter $H_0$, $\OL=1-\OM$
and $w$,  along with the nuisance parameter $\kext$, again comparing the
\ourlens\ constraints with uniform and WMAP priors, and the WMAP constraints
alone. With the WMAP data alone, $w$ is strongly degenerate with $H_0$ and
$\OL$.  Including \ourlens, which mainly provides constraints on $H_0$, the
$H_0$-$w$-$\OL$ degeneracy is partly broken.  The resulting marginalized
distribution gives $w=\wWW^{\wWWhiOne}_{\wWWloOne}$, 
consistent with a cosmological constant. The corresponding value of
Hubble's constant is $H_0 = \HWW^{\HWWhiOne}_{\HWWloOne} \kmsMpc$.

We summarize our inferences of $H_0$ and $w$ in
this variable-$w$ model in Table~\ref{tab:wConstraint}, comparing to a
similar set of alternative kinematic probes referred to in the previous section.
We see that, combining with the WMAP 5 year data set and 
marginalizing over all other parameters, the \ourlens\ data set
provides a measurement of Hubble's constant with an uncertainty of 6.9\%, with the
equation of state parameter simultaneously constrained to 18\%. 
This level of precision is better than that available from the \HST\,
KP and is competitive with the current BAO measurements.

Our results are
consistent with the results from all the other probes listed. This is not a
trivial statement: combining each data set with the WMAP 5 year prior allows us
not only to quantify the relative constraining power of each one, it also
retains the possibility of detecting inconsistencies between data sets. As it
is, it appears that all the kinematic probes listed are in agreement within
their quoted uncertainties. Some tension might 
be present if the supernovae and \ourlens\ were
considered separately from a combination of local \HST\, $H_0$
measurements and
BAO constraints, but we have no compelling reason to make such a
division. As the statistical errors associated with each probe are decreased,
other inconsistencies may arise: we might 
expect there to always be a need for 
careful pairwise data set combinations.

Finally then, we incorporate \ourlens\ into a global analysis of
cosmological data sets. As an example, we importance sample from the 
WBSw prior PDF; this is the joint posterior PDF from the joint analysis of 
WMAP5, BAO and SN data.
This prior
is already very tight, 
characterized by a median and 68\% confidence limits of
$H_0=\HWBSWprior^{\HWBSWpriorhiOne}_{\HWBSWpriorloOne} \kmsMpc$.
When we include information from
\ourlens\ with the ACS $\slope$ and OBS $\kext$ priors, we obtain
$H_0=\HWBSW^{\HWBSWhiOne}_{\HWBSWloOne} \kmsMpc$, 
a slight shift in centroid and 6\%
reduction in the confidence interval.  This is good as it shows global 
consistency in the WMAP5, BAO, SN and \ourlens\ data sets.

\begin{table*}
\begin{center}
\caption{\label{tab:wConstraint} Dark energy constraints from WMAP5 combined
with various data sets, assuming flat geometry.}
\begin{tabular}{lccccc}
\hline
 & $H_0 / \kmsMpc$ & & & $w$ & \\
\hline
WMAP5$^{\rm a,b}$              &  $74^{+15}_{-14}$             & $20\%$           & & $-1.06^{+0.41}_{-0.42}$          & $42\%$ \\
WMAP5+\HST\, KP$^{\rm a,b,c}$  & $72.1^{+7.4}_{-7.6}$          & $10\%$           & & $-1.01^{+0.23}_{-0.22}$          & $23\%$ \\
WMAP5+SN$^{\rm a,b,d}$         & $69.4^{+1.6}_{-1.7}$          & $2.3\%$          & & $-0.977^{+0.065}_{-0.064}$       & $6.5\%$ \\
WMAP5+BAO$^{\rm a,b,e}$        & $73.9^{+4.7}_{-4.8}$          & $6.6\%$          & & $-1.15^{+0.21}_{-0.22}$          & $22\%$ \\
WMAP5+Riess$^{\rm f}$          & $74.2\pm3.6^{\rm g}$          & $5.0\%$          & & $-1.12\pm0.12$                   & $12\%$ \\
{\bf WMAP5+\ourlittlelens}     & $\mathbf{\HWW^{\HWWhiOne}_{\HWWloOne}}$ & $\mathbf{6.9\%}$ & & $\mathbf{\wWW^{\wWWhiOne}_{\wWWloOne}}$ & $\mathbf{18\%}$ \\
\hline
\end{tabular}
\end{center}
The ``precisions'' in the third and fifth columns are defined   
as half the 68\% confidence interval, 
as a percentage of either 72 for~$H_0$ or -1.0 for~$w$. $\;\;$
$^{\rm a}$ \texttt{http://lambda.gsfc.nasa.gov} $\;\;$
$^{\rm b}$ \citet{KomatsuEtal09}. The $H_0$ estimate was taken from the previously listed website.$\;\;$
$^{\rm c}$ \citet{FreedmanEtal01}.$\;\;$
$^{\rm d}$ Based on the ``union'' SN samples compiled by \citet{KowalskiEtal08}.$\;\;$
$^{\rm e}$ \citet{PercivalEtal07}.$\;\;$
$^{\rm f}$ \citet{RiessEtal09}.$\;\;$
$^{\rm g}$ not marginalized over other cosmological parameters.
\end{table*}


\subsection{Future prospects}
\label{sec:H0:future}

In this paper, we have studied a single strong gravitational lens, \ourlens,
investigating in depth the various model parameter degeneracies  and
systematic effects.  At present, \ourlens\ remains the only strong lens
system with (i) time delay measurements with errors of only a few percent, and
(ii) extended source surface brightness distribution 
for accurate lens modeling; as we
have shown,  these two properties together enable the careful study and the
resulting tight constraint on $H_0$.

Table~\ref{tab:wConstraint} shows that even this one system provides
competitive accuracy on $H_0$ and $w$ for a single kinematic probe, especially
when we consider that all the other experiments involved averaging together
many independent distance measurements. What should we expect from extending
this study to many more lenses? As we showed in Section~\ref{sec:H0:nuisance},
if the data are good enough to constrain the density profile slope to a few
percent, the accuracy of the cosmological parameter inference is limited, as
it is in \ourlens, by our knowledge of the lens environment,~$\kext$. 

However, we also outlined in Section~\ref{sec:LensEnv} how using information
from numerical simulations and the photometry in the field can be used to
constrain this nuisance parameter and yield an unbiased estimate of~$H_0$.    
Furthermore, as discussed in Section~\ref{sec:H0:nuisance}, stellar 
dynamics provides significant amount of information on $\kext$ 
by limiting its permissible range of values.
While we, and also \citet{TreuEtal09} and \citet{FassnachtEtal09},
discuss how the line-of-sight contributions to~$\kext$ should average to zero
over many lens systems, lens galaxies --- like all massive galaxies ---
tend to live in locally overdense
environments, such that the local contribution to $\kext$ would be non-zero.
Careful studies of the lens environments 
(e.g. \citeauthor{MomchevaEtal06}~\citeyear{MomchevaEtal06};
\citeauthor{FassnachtEtal06}~\citeyear{FassnachtEtal06};
Blandford et al.~in preparation) and of N-body simulations with gas physics 
to determine this local contribution to $\kext$ will be crucial for
obtaining $H_0$ from a large sample of lenses. If we are able to average
together $N$ systems we should, in principle, be able to reduce our
uncertainty by~$\sqrt{N}$.  In practice, the accuracy of the combination
procedure will  sooner be limited by the systematic  uncertainty in the shape
and  centroid of the assumed $\kext$ distribution: investigating the
properties of this  distribution is perhaps the most urgent topic for further
work. Likewise, if the density profile slope cannot be constrained for each
time-delay lens individually, the details of the prior PDF assigned for~$\slope$
will become important as the ensemble grows.

In the near future, cadenced surveys such as those planned with  the Large
Synoptic Survey Telescope (LSST) and being undertaken by the 
Panoramic Survey Telescope and Rapid Response System (Pan-STARRS) will
discover large numbers of time-delay lenses, prompting us to consider
performing analyses such as the one described here on hundreds of lens
systems.  In practice, obtaining data of the quality we have presented here
for hundreds of suitable lenses will pose a significant observational
challenge. Nevertheless, \citet{DobkeEtal09}, \citet{CoeMoustakas09} and Oguri
\& Marshall (in preparation) investigate constraints on cosmological
parameters based on large samples of time-delay lenses. In particular,
\citet{CoeMoustakas09} suggest that,  in terms of raw precision and in
combination with a prior PDF from Planck, an LSST ensemble could reach
sub-percent level precision in $H_0$, and constrain $w$ to 3\% or better, 
provided that the systematic effects such as $\kext$ are under
control.  Our work has already addressed some of these systematic
effects, and will provide a basis for future analysis of large samples
of time-delay lenses and lens environment studies.


\section{Conclusions}
\label{sec:conc}

We have studied the well-observed gravitational lens \ourlens\ and used it to
infer the values of cosmological parameters; 
we outlined and followed a Bayesian approach for combining  three
data sets: \HST/ACS imaging, stellar velocity dispersion measurement, and the
time delays between the multiple images. Diagnosing the principal systematic
effects, we included two nuisance parameters ($\slope$ and $\kext$) 
into the data model to account
for them, assigning well-motivated prior PDFs and marginalizing over them.  
We draw the following conclusions:

\begin{itemize}

\item We find that the \HST/ACS images constrain the density profile slope
  parameter~$\slope = 2.08 \pm 0.03$, which we propagate through the
  cosmological parameter inference as a prior PDF. Relaxing this prior to a
  uniform distribution degrades the precision on $H_0$ from 4.4\% to 7.0\%;
  the SLACS intrinsic profile slope parameter distribution is not
  significantly more informative than the uniform prior.

\item With the ACS prior for $\slope$, we find that 
  the inferred cosmological parameters are dominated by the 
  the external convergence~$\kext$. Ray-tracing through
  the Millennium Simulation gives a PDF for~$\kext$ due to line-of-sight
  contributions that has zero mean and width~$\sim 0.04$, while using
  the galaxy number counts in the \ourlens\ field in conjunction with
  the MS gives $\kext=\kextOBS^{\kextOBShiOne}_{\kextOBSloOne}$.
  
\end{itemize}

Using our most informative priors on the two nuisance parameters, we 
arrive at the following cosmographic inferences:

\begin{itemize}

\item In the K03 cosmology ($\OM=0.3$, $\OL=0.7$, $w=-1$, and uniform
  $H_0$), we obtain from the \ourlens\ data set $H_0 = \HK \pm \HKerrOne
  \kmsMpc$ (68\% CL).  The $4.4\%$ error includes both statistical
  and dominant systematic uncertainties, through the marginalization described
  above.  This is a significant improvement to
  the earlier measurement of $H_0=75^{+7}_{-6}\kmsMpc$ by
  \citet{KoopmansEtal03}.

\item Time-delay lenses are sensitive primarily to $H_0$ but are
  weakly dependent on other cosmological parameters; the
  lensing measurement of $H_0$ is robust and useful
  for studying dark energy when combined with other
  cosmological probes. We find that for \ourlens\ the cosmographic
  information 
  can be summarized as a shifted log normal
  probability distribution for the time-delay distance~$\tdist$ in
  units of Mpc, with the three parameters $\lambda_{\rm D}=\Dtfitlam$,
  $\mu_{\rm D}=\Dtfitmu$ and $\sigma_{\rm D}=\Dtfitsig$. 

\item In a $\Lambda$-CDM cosmology (with $w=-1$), the \ourlens\ data set
  breaks the degeneracy between $\OM$ and $\OL$ in the WMAP 5 year data set, 
  and constrains the curvature parameter to be zero to $2.0\%$ (95\% CL), 
  a level of
  precision similar to those afforded by the current Type Ia SNe sample.

\item \ourlens\ in combination with the WMAP 5 year data set, 
  assuming flatness and
  allowing (a time-independent) $w$ to vary, 
  gives $H_0 = \HWW^{\HWWhiOne}_{\HWWloOne} \kmsMpc$ and
  $w=\wWW^{\wWWhiOne}_{\wWWloOne}$ (68\% CL).  
  
  These are significant improvements
  to the WMAP5 only constraints of 
  $H_0=\HWWprior^{\HWWpriorhiOne}_{\HWWpriorloOne}\,\kmsMpc$ and
  $w=\wWWprior^{\wWWpriorhiOne}_{\wWWpriorloOne}$.
  \ourlens\ is as competitive as the current BAO data in determining
  $w$ when combined with WMAP5.

\end{itemize}

Our detailed analysis of \ourlens\ provides the framework for using large
samples of time-delay lenses as cosmological probes in the near future.  We
anticipate the local contribution to $\kext$, which would not average away
with a large sample of lenses, being  the dominant residual systematic
error.   Several lens environment studies to circumvent this are underway;
with the effects from $\kext$ accurately modeled, future samples of time-delay
gravitational lenses should be a competitive cosmological probe.

\acknowledgments We thank M.~Brada{\v c}, J.~Hartlap, E.~Komatsu, J.~P.~McKean and
P.~Schneider for useful discussions.  We are grateful to the anonymous
referee whose suggestions and comments helped clarify parts of the paper.
S.H.S. is supported in part through the Deutsche 
Forschungsgemeinschaft under the project SCHN 342/7--1.  
C.D.F. acknowledge support under the \HST\
program \#GO-10158. Support for program \#GO-10158 was provided by
NASA through a grant from the Space Telescope Science Institute, which
is operated by the Association of Universities for Research in
Astronomy, Inc., under NASA contract NAS 5-26555.  C.D.F.
acknowledge the support from the European Community's Sixth Framework
Marie Curie Research Training Network Programme, contract no.
MRTN-CT-2004-505183 ``ANGLES.''  R.D.B. acknowledges support through NSF 
grant AST 05-07732. L.V.E.K. is supported in part through an
NWO-VIDI career grant (project number 639.042.505).  T.T. acknowledges
support from the NSF through CAREER award NSF-0642621, by the Sloan
Foundation through a Sloan Research Fellowship, and by the Packard
Foundation through a Packard Fellowship.  This work was supported in
part by the NSF under award AST-0444059, the TABASGO foundation in the
form of a research fellowship (P.J.M.), and by the US Department of
Energy under contract number DE-AC02-76SF00515.  Based in part on
observations made with the NASA/ESA \textit{Hubble Space Telescope},
obtained at the Space Telescope Science Institute, which is operated
by the Association of Universities for Research in Astronomy, Inc.,
under NASA contract NAS 5-26555. These observations are associated
with program \#GO-10158.


\bibliographystyle{apj}
\bibliography{ms}


\appendix


\section{Probability Theory for Measuring Cosmological Parameters}
\label{app:H0ProbTheory}

In this appendix, we describe how we derive the expressions for the
likelihoods of the time delay and the ACS data sets stated in Section
\ref{sec:H0ProbTheory:Like}.  We also provide the details on the
sampling techniques for calculating the posterior probability density
of cosmological parameters.

\subsection{Simplification of the likelihoods}
\label{app:H0ProbTheory:ParamConstraint:Simp}

For \ourlens, we can simplify the marginalization of the likelihoods
$P(\tdelayVec|\pars)$ and $P(\dataVec|\pars)$ in Equation
(\ref{eq:CosmoparsPosterior:sec}) based on the following facts, which
are either from Paper I or shown in Section \ref{sec:LensModel}:
\begin{itemize}
\item from Paper I, the top $\imModelSet$ models led to equal evidence values
  $P(\dataVec | \slope, \lenspars, \dpsiMPVec, \imModelSet)$ (within the
  uncertainties).
\item in Section \ref{sec:LensModel}, the likelihood function
  $P(\tdelayVec|\pars)$ is approximately constant for these top $\imModelSet$
  models for a given cosmology.
\item in Section \ref{sec:LensModel}, for good data models $\imModelSet$ of
  \ourlens, the potential corrections $\dpsiVec$ do not change significantly
  the predicted values of the Fermat potential, i.e., the simply-parametrized
  SPLE initial model provides an unbiased estimator for the Fermat potential.
\item the likelihood $P(\vdisp|\pars)$ is also constant for the various
  $\imModelSet$ because the dynamics modeling is independent of the lensed
  image processing models $\imModelSet$.
\item simulations suggest that the potential corrections $\dpsiVec$ are
  sharply peaked about the most probable values $\dpsiMPVec$.
\end{itemize}

With the above results, the Fermat potential can be more easily
computed from the SPLE model: there is a strong correlation between
the $\fpdiff$ and $\slope$, and we obtain the relation between
$\fpdiff$ and $\slope$ by evaluating them at several discrete $\slope$
values and interpolating between them.  For notational simplicity, we
consequently drop the nearly-true independences of $\fpdiff$ on
$\lenspars$ and $\dpsiVec$.  In computing the predicted $\fpdiff$, the
average source position of the four mapped (via the lens equation)
image positions on the source plane is used. 
Denoting the dependence of $\fpdiff$ on
$\slope$ for $\kext=0$ and a given $\imModelSet$ as
$q(\slope,\imModelSet)$ and using the mass-sheet degeneracy relation
for the dependence of $\fpdiff$ on $\kext$, we obtain Equations
(\ref{eq:fprelation:sec}) through (\ref{eq:tdelayLikeSimp:sec}) for
the likelihood of the time delay data.\footnote{The Fermat potential
  is independent of cosmological parameters because we work in terms
  of the scaled lens potential in angular (arcsecond) units.
  Cosmological parameters and redshifts are only needed for deriving
  physical quantities of the lens system, such as the one-dimensional
  velocity dispersion (calculable from the Einstein radius) of the
  lens, mass of the lens, and the physical extent of the lens/source.}

For the likelihood of the ACS data, we work with the SPLE potential
model on the basis that the Fermat potential is insensitive to the
potential corrections and the potential corrections only slightly
alter the ranking of $\imModelSet$ models.  We can choose the
$\imModelSet$ to be one of the top models, say $\imModelFive$ (since
the normalization in $P(\cosmopars | \tdelayVec, \dataVec, \vdisp)$ is
irrelevant), and drop the dependence on $\dpsiVec$ in the likelihood
of the ACS data to simplify part of the integrand in Equation
(\ref{eq:CosmoparsPosterior:sec}):
\bea
 & & \int {\rm d}\dpsiVec\ {\rm d}\srVec\ {\rm d}\imModelSet\ P(\dataVec | \slope, \lenspars, \dpsiVec, \srVec, \imModelSet)\cdot\nonumber\\
& & \ \ \ \ \ \  P(\srVec | \lambda, \regSet) P(\imModelSet) P(\dpsiVec)\nonumber \\
\label{eq:CosmoparsPosteriorSimp1}
& \propto  & \sim \int {\rm d}\srVec\ P(\dataVec | \slope, \lenspars,
\srVec, \imModelSet=\imModelFive) P(\srVec | \lambda, \regSet)
\nonumber \\ 
\label{eq:CosmoparsPosteriorSimp2}
& = & P(\dataVec | \slope, \lenspars, \imModelSet=\imModelFive), 
\eea
which is also Equation (\ref{eq:CosmoparsPosteriorSimp1:sec}).  In
deriving the above equation, we assume that the representative set of
models $\imModelSet$ obtained in \citet{SuyuEtal09} are equally
probable a priori (i.e., the prior $P(\imModelSet)$ is constant).  For
the priors $P(\srVec | \lambda, \regSet)$ and $P(\dpsiVec)$, we use
quadratic forms of the regularizing function.  Specifically, we try
zeroth-order, gradient and curvature forms for $P(\srVec | \lambda,
\regSet)$, and the curvature form for $P(\dpsiVec)$, as described in
\citet{SuyuEtal06} and \citet{SuyuEtal09}.  As a reminder, the
quantity $P(\dataVec | \slope, \lenspars, \imModelSet=\imModelFive)$
is the Bayesian evidence from source reconstruction given the lens
model parameters \{$\slope$, $\lenspars$\} and the data model
$\imModelFive$; this evidence value is calculable based on
\citet{SuyuEtal09}.

\subsection{Importance sampling}
\label{app:H0ProbTheory:ImpSamp}

In practice, we can incorporate the various likelihoods in Equation
(\ref{eq:CosmoparsPosteriorFullSimp2:sec}) by importance sampling 
the prior distribution \citep[see e.g.\ ][for an
introduction]{LewisBridle02}. This is a method for calculating integrals over
a PDF $P_2$ when all we have is samples drawn from some other PDF
$P_1$. Consider the expectation value of a parameter $x$:
\bea
\langle x \rangle_2 &=& \int x \cdot P_2(x)\, dx, \\
                    &=& \int x \frac{P_2(x)}{P_1(x)} \cdot P_1(x)\, dx.
\eea
The process of weighting the samples from $P_1$ by the ratio $P_2(x)/P_1(x)$
is called importance sampling. It works most efficiently 
when $P_1$ and $P_2$ are quite similar, and fails if $P_1$ is zero-valued over
some of the range of $P_2$, or if the sampling of $P_1$ is
too sparse. 

In our case, we would like to calculate integrals over, for example,
$P_2 = P(\cosmopars,\slope,\kext,\aniso |\tdelayVec,\dataVec,\vdisp)$,
while the prior is written simply $P_1 =
P(\cosmopars,\slope,\kext,\aniso |\dataVec)$ (recall that $\dataVec$
is used to provide a prior on $\slope$).  Using Bayes' theorem, we can
write
\bea 
\lefteqn{P(\cosmopars,\slope,\kext,\aniso |\tdelayVec,\dataVec,\vdisp) \propto} \nonumber \\
&& \ \ \ \ P(\tdelayVec,\vdisp| \cosmopars,\slope,\kext,\aniso) P(\cosmopars,\slope,\kext,\aniso|\dataVec),\nonumber\\
\lefteqn{{\rm i.e.}\;\;\; P_2 \propto P(\tdelayVec,\vdisp| \cosmopars,\slope,\kext,\aniso) P_1.}
\eea
From this we can see that the weight we must attach to each sample
from the prior is just the value of the likelihood
$P(\tdelayVec,\vdisp| \cosmopars,\slope,\kext,\aniso)$.  Note that these
weights can be rescaled by an arbitrary factor, which can be important
in retaining numerical stability.

We apply this technique to perform the
marginalization in Equation
(\ref{eq:CosmoparsPosteriorFullSimp2:sec}).  Specifically, we have samples of
$P(\cosmopars)$, $P(\slope|\dataVec,\imModelSet=\imModelFive)$,
$P(\kext)$ and $P(\aniso)$, and employ importance sampling to obtain
$P(\cosmopars|\tdelayVec,\dataVec,\vdisp)$.


\end{document}

%% file: ms_macro.tex
\usepackage{latexsym}
\usepackage{amssymb}
\usepackage{graphicx}

\newcommand{\bd}{\begin{displaymath}}
\newcommand{\ed}{\end{displaymath}}
\newcommand{\be}{\begin{equation}}
\newcommand{\ee}{\end{equation}}
\newcommand{\beaa}{\begin{eqnarray*}}
\newcommand{\eeaa}{\end{eqnarray*}}
\newcommand{\bea}{\begin{eqnarray}}
\newcommand{\eea}{\end{eqnarray}}

\def\ourlens{B1608$+$656{}}
\def\ourlittlelens{B1608{}}
\def\paperI{Paper~I{}}

\newcommand{\boldsymbol}[1]{\mbox{\boldmath{${#1}$}}}

\newcommand{\bmath}{\vec}

\def\dataVec{\boldsymbol{d}}

\def\regSet{\boldsymbol{\mathsf{g}}}

\def\srVec{\boldsymbol{s}}

\def\imCM{\boldsymbol{\mathsf{C}}_{\mathrm{D}}}

\def\dpsiVec{\boldsymbol{\delta \psi}}
\def\dpsiMPVec{\boldsymbol{\delta \psi}_{\mathrm{MP}}}

\def\HST{{\it HST}{}}
\def\blurSet{\boldsymbol{\mathsf{B}}}

\def\dustSet{\boldsymbol{\mathsf{K}}}

\def\glightVec{\boldsymbol{l}}

\def\vdisp{\sigma}
\def\slope{\gamma'}
\def\aniso{r_{\rm ani}}

\def\lenspars{\boldsymbol{\eta}}

\def\Rein{R_{\rm{Ein}}}

\def\tdelayVec{\boldsymbol{\Delta t}}
\def\tdelay{\Delta t}

\def\imModelSet{\boldsymbol{M}_{\rm D}}
\def\imModelFive{\boldsymbol{M}_{5}}
\def\imModelNum{\boldsymbol{M}}

\def\psiVec{\boldsymbol{\psi}}

\def\psiI{\psi_0}
\def\psiIVec{\boldsymbol{\psi_0}}
\def\kext{\kappa_{\rm ext}}

\def\fp{\phi}
\def\fpdiff{\Delta\phi}

\def\tdist{D_{\rm \Delta t}}

\def\zd{z_{\rm d}}
\def\zs{z_{\rm s}}

\def\kmsMpc{\rm{\, km\, s^{-1}\, Mpc^{-1}}}
\def\kms{\rm{\, km\, s^{-1}}}

\def\OM{\Omega_{\rm m}}
\def\OL{\Omega_{\rm \Lambda}}
\def\Ok{\Omega_{\rm k}}
\def\pars{\boldsymbol{\xi}}
\def\cosmopars{\boldsymbol{\pi}}

\def\kextOBS{0.10}
\def\kextOBShiOne{+0.08}
\def\kextOBSloOne{-0.05}

\def\DtOne{(5.16^{+0.29}_{-0.24})\times10^3}
\def\Dtfitlam{4000.}
\def\Dtfitmu{7.053}
\def\Dtfitsig{0.2282}

\def\HK{70.6}

\def\HKerrOne{3.1}

\def\HKnodyn{68.1}
\def\HKnodynhiOne{+3.7}
\def\HKnodynloOne{-6.4}

\def\HWW{69.7}
\def\HWWhiOne{+4.9}
\def\HWWloOne{-5.0}

\def\HWWprior{74}
\def\HWWpriorhiOne{+15}
\def\HWWpriorloOne{-14}

\def\wWW{-0.94}
\def\wWWhiOne{+0.17}
\def\wWWloOne{-0.19}

\def\wWWprior{-1.06}
\def\wWWpriorhiOne{+0.41}
\def\wWWpriorloOne{-0.42}

\def\OkWO{-0.005}
\def\OkWOhiTwo{+0.014}
\def\OkWOloTwo{-0.026}
\def\OkWOTwo{-0.031<\Ok<0.009}

\def\HWBSW{70.4}
\def\HWBSWhiOne{+1.5}
\def\HWBSWloOne{-1.4}

\def\HWBSWprior{70.3}
\def\HWBSWpriorhiOne{+1.6}
\def\HWBSWpriorloOne{-1.5}